\documentclass[onecolumn,usenatbib]{mn2e}
\usepackage{natbib}
\usepackage{amsmath}
\usepackage{graphicx}
\usepackage{amsbsy}
\usepackage{epsf}
\def\X{{\mathrm{x}}}
\def\Y{{\mathrm{y}}}

\def\n{{\rm n}}

\def\p{{\rm p}}

\def\l{\Lambda}
\def\c{{\rm c}}
\def\s{\Sigma}

\newcommand{\be}{\begin{equation}}
\newcommand{\ee}{\end{equation}}
\newcommand{\beq}{\begin{eqnarray}}
\newcommand{\eeq}{\end{eqnarray}}
\newcommand{\bear}{\begin{eqnarray}}
\newcommand{\eear}{\end{eqnarray}}
\newcommand{\ba}{\begin{array}}
\newcommand{\ea}{\end{array}}
\begin{document}

\title{Superfluid hyperon bulk viscosity and the r-mode instability of rotating neutron stars}

\author [Haskell \& Andersson]{B. Haskell, N. Andersson\\
School of Mathematics, University of
Southampton, Southampton SO17 1BJ, UK}
\maketitle
\begin{abstract}
In order to establish whether the unstable r-modes in a rotating neutron star  provide a detectable source of gravitational waves, we need to understand the details of the many dissipative processes that tend to
counteract the instability. It has been established that the bulk viscosity due to exotic particles, like hyperons, may be particularly important in this respect. However, the effects of
hyperon superfluidity have so far not been fully accounted for. While the associated suppression of the reaction rates that give rise to the bulk viscosity has been estimated, superfluid aspects of the fluid dynamics have not been considered. In this paper we determine the r-mode instability window for a neutron star with a $\Sigma^{-}$ hyperon core, using the appropriate multifluid formalism  including, for the first time, the effect of the ``superfluid'' bulk viscosity coefficients. We demonstrate that, even though the extra terms may increase the bulk viscosity damping somewhat, their
presence does not  affect the qualitative features of the r-mode instability window.

\end{abstract}


\section{Introduction}

Comprising one and a half solar masses  inside a radius
of roughly ten kilometers, neutron stars provide an arena where many
extremes of physics meet. A detailed model of neutron star dynamics
must account for  strong magnetic fields, various
superfluid/superconducting components, the interaction between
the crust nuclei and the fluid, as well as
exotic states of matter that may be present in the neutron star
core. Needless to say, it is a formidable task
to construct such a model.
Especially since it requires an understanding of physics well beyond
the laboratory. While the equation of state for matter
approaching the nuclear saturation density  $n_0 \approx 0.16$~fm$^{-3}$
(corresponding to $2.48\times 10^{14}$~g/cm$^3$) is 
quite well understood, it seems unlikely that laboratory
experiments will ever be able to probe the densities expected
in the deep core of a neutron star (above several times $n_0$).

Despite decades of research into the supranuclear equation of
state, considerable uncertainties remain. Furthermore, neutron star (NS)
observations have only recently begun to reach the level of precision necessary to constrain the theoretical
models in a severe way. Very recent results by \citet{Ozel} suggest that current measurements of NS masses and radii can be used to rule out several nuclear equations of state and support the notion that the core of a NS should contain exotic particles, such as pions, kaons, hyperons or even a deconfined quark condensate.
Establishing observable signatures of the presence of such exotic states of matter
is a priority for modelling in this area.

During the last decade, the notion that gravitational waves (GWs) may
drive the so-called r-modes of a rotating neutron star unstable,
and that this may lead to the star spinning down on a timescale
of weeks to months, has been discussed in a number of papers, see \citet{review} and \citet{anderssonREVIEW} for reviews.
The r-mode instability initially
attracted attention because  it provided
a mechanism that could spin a newly born neutron star down dramatically, releasing GWs at a level that might be detectable in the process \citep{O98}.
For the purpose of GW detection, rapidly rotating accreting neutron stars in
low-mass X-ray binaries (LMXBs)
have also attracted attention. In these systems, the r-modes could provide a mechanism for torque balance \citep{Bildsten,AKS} and, in some cases, lead to persistent GW emission \citep{strange, wagoner2004, Nayyar}.
This scenario currently holds interesting prospects for detection, albeit with considerable technical difficulties \citep{watts2008}.

\begin{figure}
\centerline{\includegraphics[height=7cm,clip]{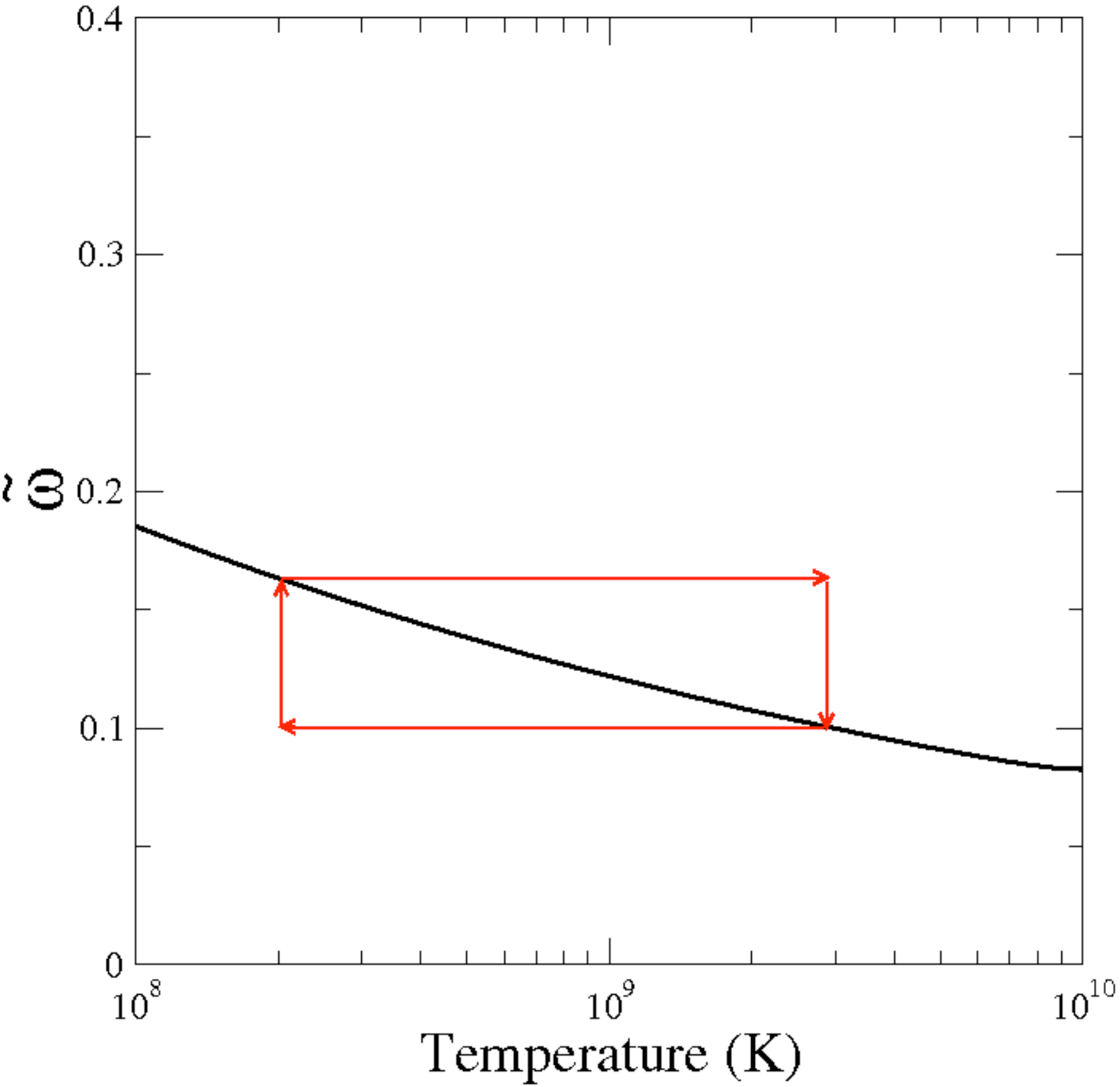}\includegraphics[height=7cm,clip]{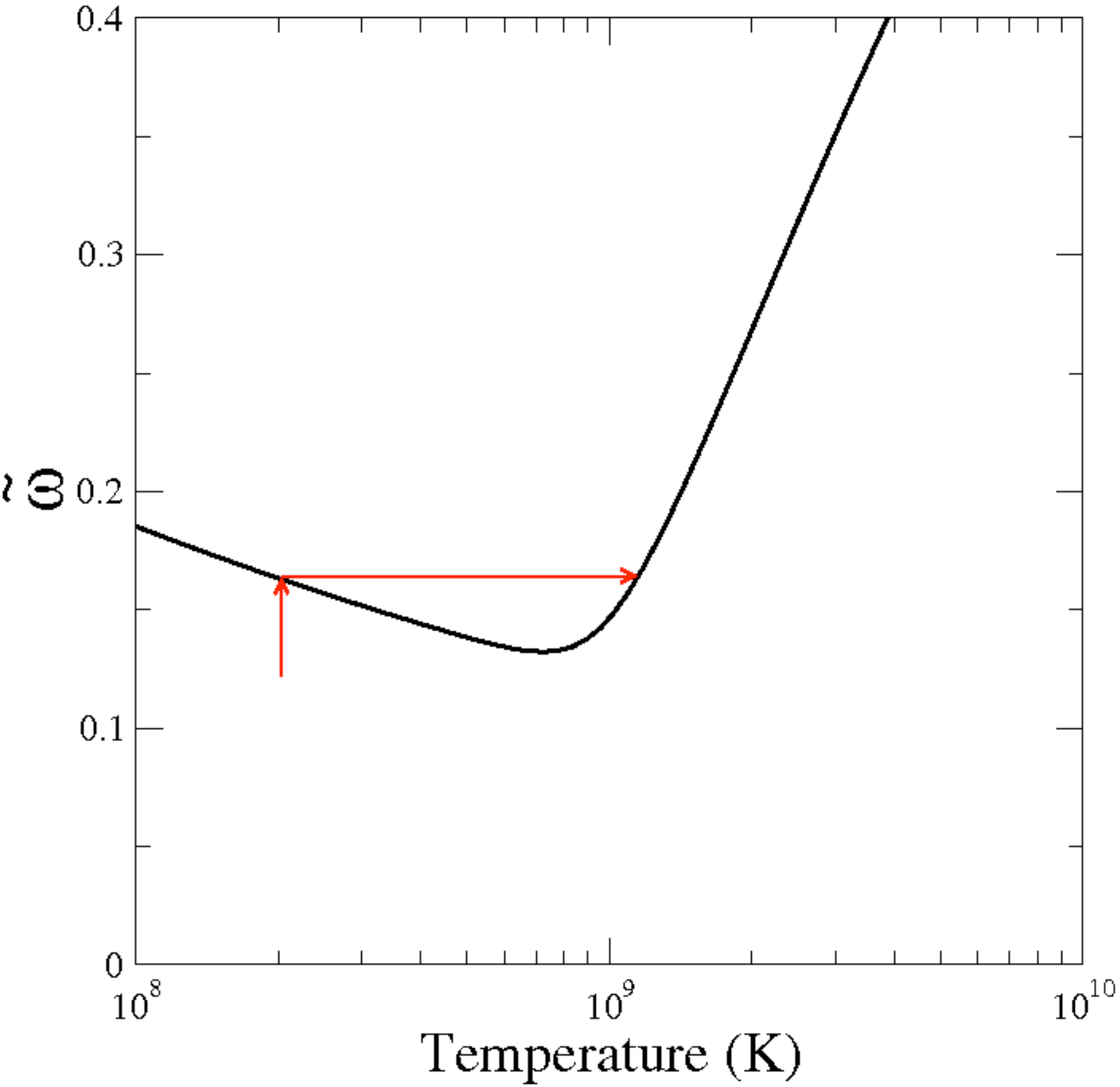}}
\caption{The r-mode instability window for stars with $M=1.4 M_{\odot}$  and $R=13$km. On the left we have a star where the dominant damping mechanism at low temperatures is shear due to electrons in an Ekman layer at the base of the crust, on the right  a star where the main damping mechanism above $10^9$ K is bulk viscosity due to hyperons. The rotation rate is expressed in terms of the parameter $\tilde{\omega}=\omega/\sqrt{G M/R^3}$. The instability curve in the left panel always has a negative slope and the system undergoes a limit cycle as described in the text. This is in contrast with the instability curve on the right which has a positive slope in the $10^9$ K region, which could halt the thermal run-away and lead to persistent emission of gravitational waves. }
\label{windows}
\end{figure}

Not surprisingly, the details of the supranuclear equation of state (EOS) are key to understanding the r-mode instability. In a real neutron star many (viscous) mechanisms compete with the GW driving of the r-mode. If the star contains exotic particles, such as hyperons, additional dissipation channels may become relevant. At first sight, it would appear that additional damping should reduce the chances of the r-mode GWs being detectable as it would reduce the region of parameter space where the instability is active. However, this is not necessarily  the case. As an illustration of this, let us
consider the instability in the temperature range $10^7-10^{10}$~K. First assume that the main damping mechanism is due to a viscous Ekman layer at the core-crust interface (for a discussion see \citet{eck1, eck2})  (the left panel of figure \ref{windows}). Standard bulk viscosity due to modified URCA reactions is only relevant above $10^{10}$ K and is not included in this example (we shall discuss this in detail in section \ref{rmode}). In this case a NS in an LMXB, which will heat up to a core temperature of a few times $10^8$~K, would spin up due to accretion and enter the instability window before the star reaches the break-up limit. The shear from the unstable r-mode then heats the star  and the emission of GWs spins the star down until it returns to the stable region. At this point the star will cool down and the cycle can begin again \citep{levin1}. Unfortunately, with the current estimates for the nonlinear saturation amplitude \citep{Arras,Brink} the duty cycle for this scenario is very low, meaning that the star would emit brief bursts of gravitational radiation and we would observe most systems in quiescence. Let us contrast this model with a system
 where we have added the effect of hyperon bulk viscosity (the right panel of figure \ref{windows}). In this case, extra damping leads to a positive slope of the curve in the $10^9$ K region and there are now
three possible scenarios. Depending on the mode amplitude and on the exact details of the damping, the star could either i) execute the cycle we have already described, or ii) the heating may be sufficient for the
system to evolve horizontally all they way to the positively sloped part of the curve before GW emission has time to spin the star down \citep{Bondarescu}. Finally, it may be the case that, iii) the heating due to accretion is such that the system becomes unstable in the region with positive slope. In the last two scenarios the system will not be able to evolve away (significantly) from the instability curve, and should become a persistent source of GWs \citep{strange}. This hypothesis was  examined by \citet{Nayyar}, who found that, in the case of a NS with a hyperon core, persistent emission is possible over a wide range of parameters for the bulk viscosity and the EOS.

LMXBs contain old NSs which will have cooled well below the temperature at which the hyperons (and other components of the star, such as neutrons and protons) are likely to become superfluid.
Superfluidity adds dimensions to the problem as, not only does it reduce the reaction rates for hyperon creation processes, it also increases the dynamical degrees of freedom of the system. In general, we have to work with multifluid hydrodynamics, with each particle species (potentially) leading to an independent flow. This leads to the appearance of new families of oscillation modes \citep{Epstein, LM94, ac2001} and also has profound consequences for viscous dissipation. Not only are there new dissipation mechanisms, such as mutual friction between the components \citep{ALS, Ma,Mb,Trev}, but bulk- and shear viscosity can no longer be described by single coefficients. In general, there are many additional viscosity coefficients. In the context of neutron stars, this was first pointed out by \citet{CQG}. The particular case of a hyperon core was first considered by \citet{Gusakov2008} and has recently been analysed in detail by \citet{prep}. In the simplest case, that of a core comprising neutrons, protons, electrons and $\Sigma^{-}$ hyperons,  one can show that the problem is very similar to that for superfluid Helium \citep{helium} and can be described by three bulk viscosity and one shear viscosity coefficient. One would, of course, also need to account for mutual friction between the various superfluid components. It has, however, been shown in several calculations that in a NS composed of neutrons, protons and electrons, mutual friction is unlikely to have a significant effect on the r-mode instability \citep{LM,LY,rmode}. As the nature of mutual friction involving hyperons is largely unknown, we shall assume that it can also be neglected (the veracity of this assertion obviously needs to be
checked by detailed work in the future). We will focus on the bulk viscosity, which is expected to give the main contribution
to the r-mode damping \citep{PBJones,LO,haensel,Nayyar}.  Although the relevance of the new ``superfluid'' bulk viscosity coefficients is well established for superfluid Helium, their effect has mostly been neglected in the study of NS oscillations.
The notable exception is the work of \citet{Gusakovsound} who studied the damping of sound waves in a dense superfluid hyperon core and showed that the additional bulk viscosity terms can play a significant role.

It is clearly important to understand the role of the extra damping coefficients and refine our theoretical understanding of the bulk viscosity, as we have seen that the nature of the damping mechanisms can have profound consequences for the r-mode instability and the associated GW emission. In fact, direct GW detection from these systems should allow us to discern  whether the star is emitting persistently or is executing a limit cycle. This would provide valuable information on the physics of the NS interior.
The purpose of this paper is  to study the effect of superfluid hyperon bulk viscosity on the r-mode instability window. The formalism for studying r-modes in multifluid neutron stars has been developed by \citet{rmode} and we shall extend it to include hyperon bulk viscosity, thus considering for the first time the global dynamics of a multifluid NS with a hyperon core.

\section{Size of the hyperon core}

\begin{figure}
\centering
\includegraphics[height=6cm,clip]{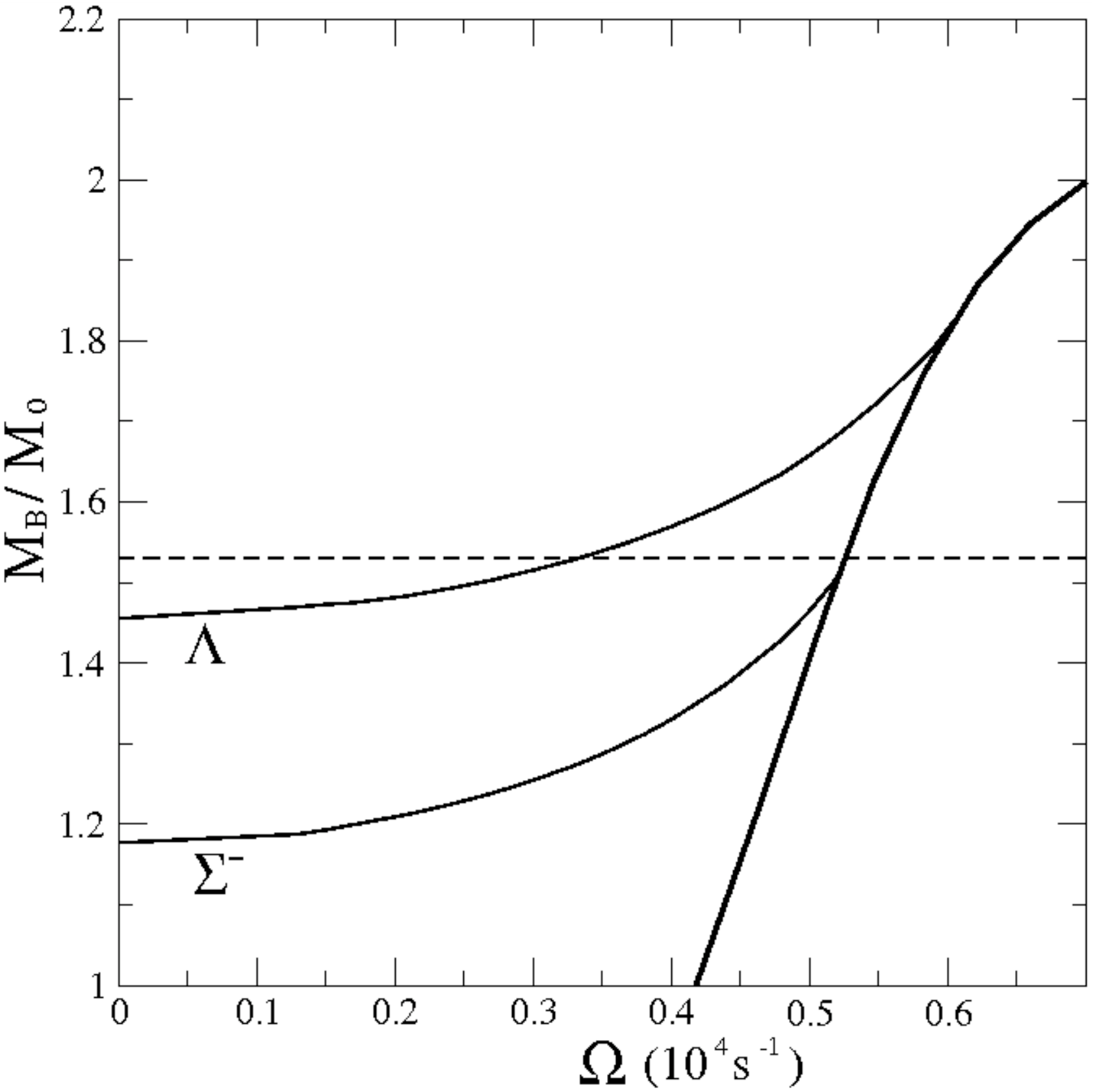}
\includegraphics[height=6cm,clip]{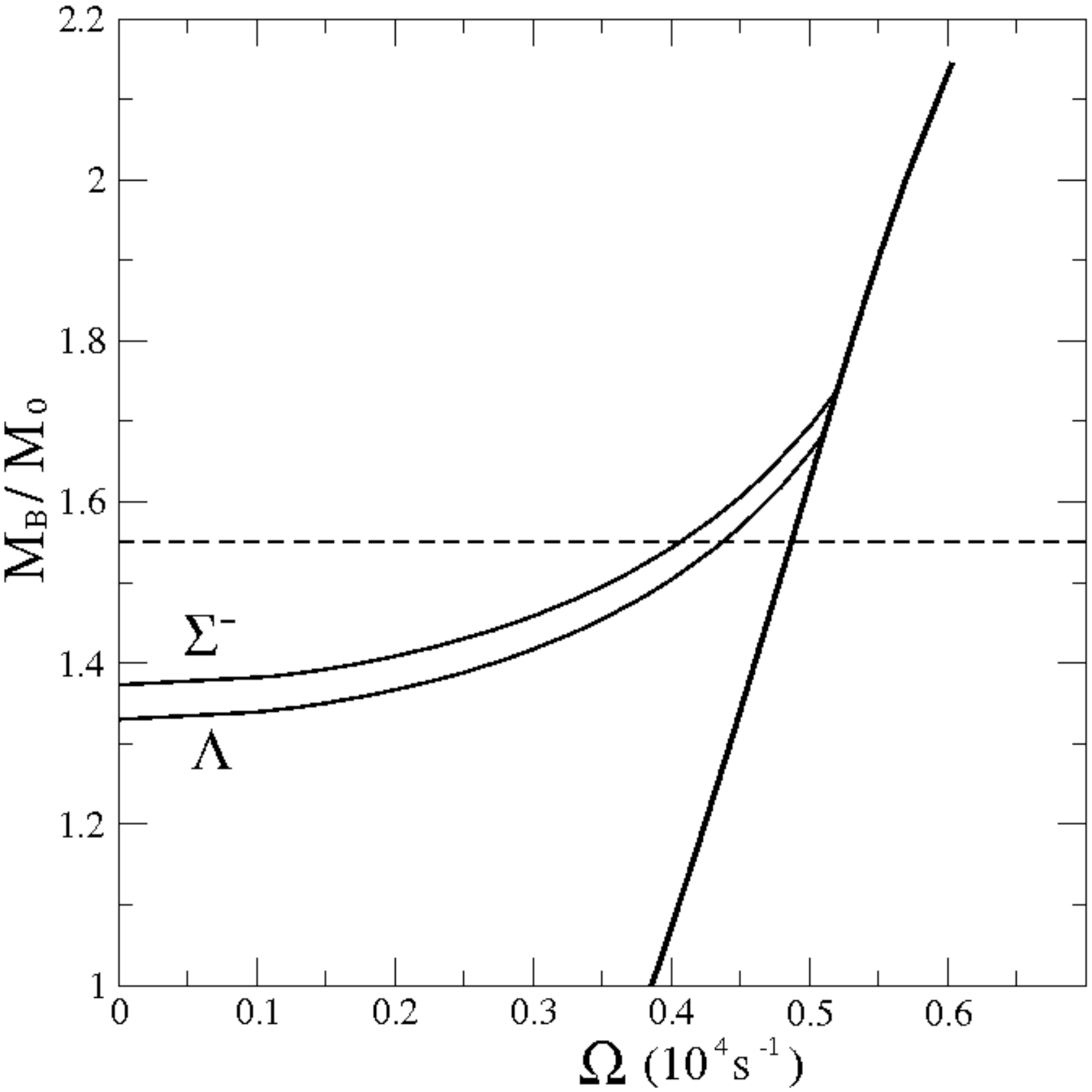}
\caption{We plot the rotation rate at which hyperons and deconfined
quarks first appear for neutron stars of different baryon masses ($M_B$).
The data was obtained using the EOS $G_{240}$ \citep{Glenbook} (left panel) and case III of \citet{Glen}
(right panel). For both equations of state stars with masses just above the threshold for the appearance of hyperons would have their exotic core spun out (or significantly reduced) as they approach the Keplerian rotation frequency. The horizontal line indicates the baryon mass that corresponds to a gravitational mass of $M=1.4M_\odot$ in a non-rotating model.}
\label{rot1a}
\end{figure}

Let us begin by  considering the extent of the hyperon-rich core. It is well known that the central density of a star decreases as the rotation rate increases. This means that the threshold density for the presence of  hyperons  moves closer to the centre of the star and it could, in theory, be possible to ``spin out'' such an exotic core completely.
Examples of this effect are shown in Figure~\ref{rot1a}. The figure shows how, for two relativistic
mean field EOS, the size of the hyperon core (in the equatorial plane) depends on the star's rotation.
Let us focus on stars with a constant baryon mass (horizontal lines in the figure) corresponding to a gravitational
mass of $1.4M_\odot$ in the non-rotating limit. Then,
in the case of the $G_{240}$ EOS \citep{Glenbook} (the left panel), there are no $\Lambda$ hyperons present when the break-up limit
is reached and the $\Sigma^-$ core is very small. In the other example, for the EOS given by case 3 in \citet{Glen}, both
the $\Lambda$ and the $\Sigma^-$ hyperons are completely spun out before the break-up limit is achieved.

\begin{figure}
\centering
\includegraphics[height=5cm,clip]{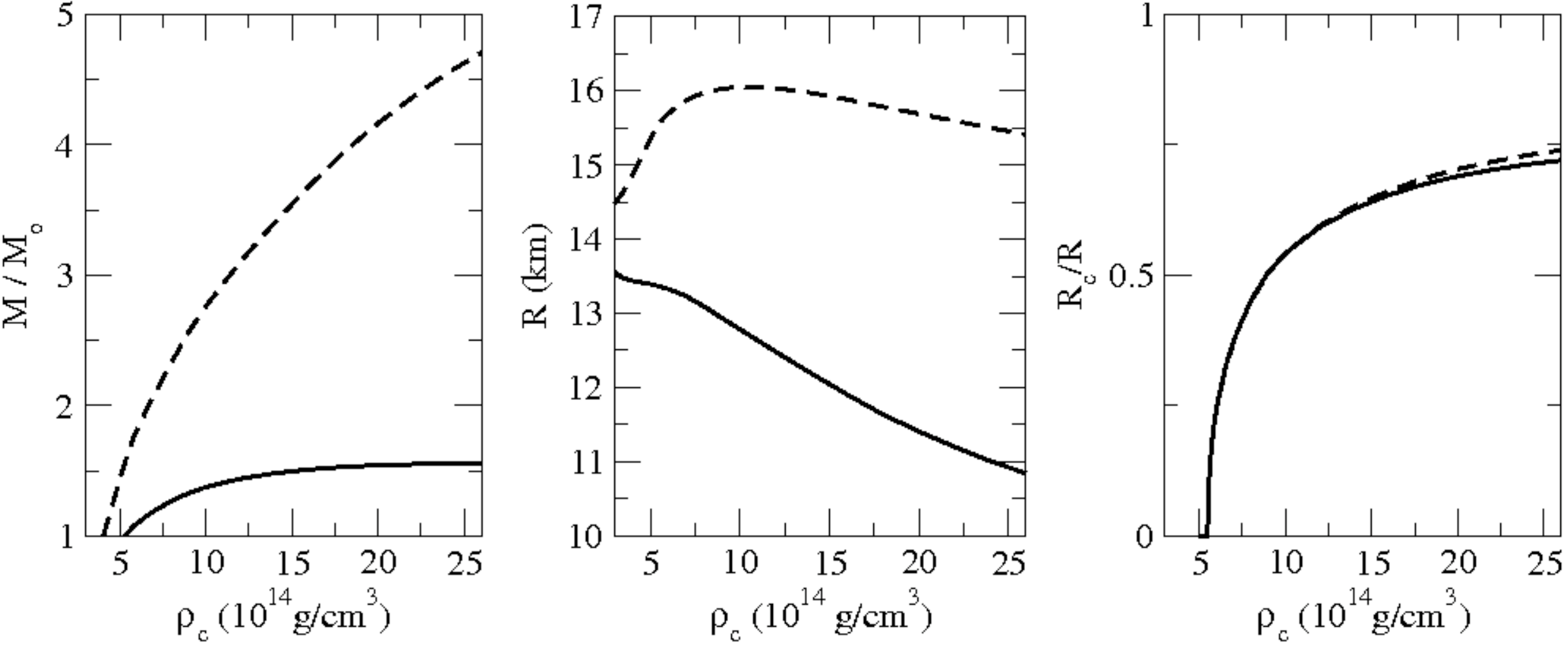}
\caption{Comparing results for neutron star models calculated from the
EOS $G_{240}$ in both Newtonian gravity (dashed lines)
and full General Relativity (solid lines). The first
two panels illustrate the well-known fact that the two theories lead to
rather different results for neutron star masses and radii given
the same central density. From these panels it is clear why it is not
particularly meaningful to use a realistic equation of state in a
Newtonian study. The third panel is relevant for studies of effects
due to the presence of a hyperon core. It is shown how the ratio of the
hyperon core (here $R_c$ corresponds to the radius at which the density is
such that the $\Sigma^-$ hyperon first appears) to the radius of the star
varies with the central density. Interestingly, these results indicate that,
despite the stellar models being very different, the relative size of the core is almost the
same in Newtonian theory and General Relativity.}
\label{grvsn}
\end{figure}

It is obviously relevant to quantify this effect and consider its impact on the r-mode instability window.
Before doing this it is, however, important to discuss how reliable Newtonian estimates of the size of an exotic
core are. This is relevant since r-mode viscous damping timescales
are mostly calculated in Newtonian theory so far (although \citet{Nayyar} calculate the background stellar model in general relativity and \citet{jose} calculate shear viscosity damping for relativistic r-modes).
A useful answer is provided in Figure~\ref{grvsn}. The results demonstrate that
the \underline{relative size} of the core is  well approximated by the Newtonian model,
even though the actual size of the star is off by a large amount compared to the relativistic result.

\begin{figure}
\centerline{\includegraphics[height=7cm,clip]{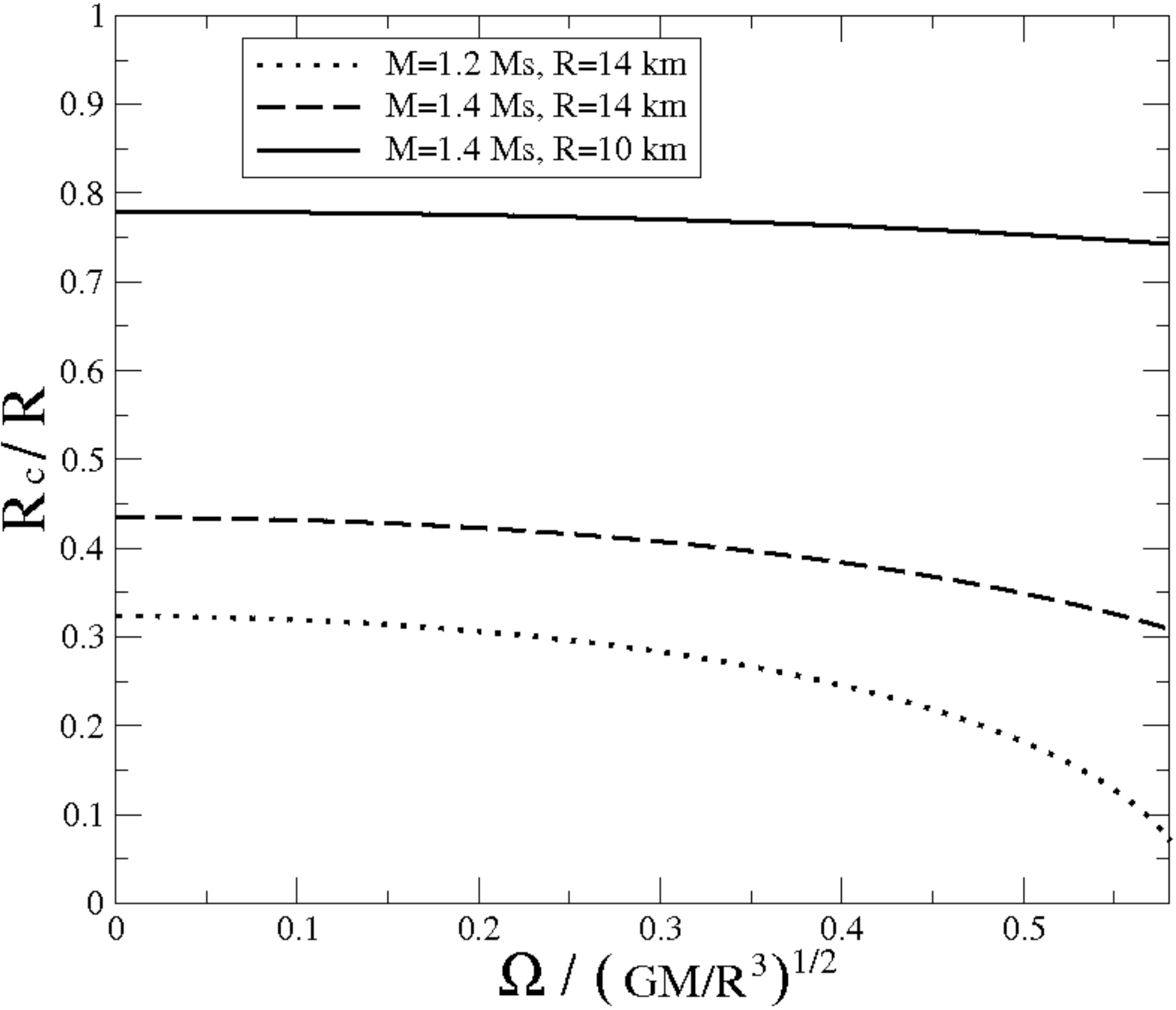}}
\caption{Extent of the exotic core, for two models with $M=1.4 M_{\odot}$, for $R=14$ km and $R=10$ km and a model with $M=1.2 M_{\odot}$ and $R=14$ km (the equation of state is taken to be an $n=1$ polytrope). The graph extends to the Keplerian breakup frequency. For the stellar model with $M=1.4 M_{\odot}$ and  $R=14$ km the extent of exotic core is reduced at the breakup frequency, while for the model with $M=1.2 M_{\odot}$ the hyperon core is essentially spun out before the breakup rotation rate is reached.}
\label{radius}
\end{figure}

This suggests that, even though it is generally not meaningful to use realistic EOS in Newtonian models
(since the radius of a star with a given mass/dentral density differs so much from the relativistic model),
one can study the relative size of the hyperon core also in Newtonian theory. To do this, we construct a sequence of stars spinning at different rotation rates but with the same total mass.
For simplicity, we now restrict ourselves to rotating $n=1$ polytropes (a useful model since both the background and the r-mode solution can be studied analytically).
Following the analysis in \citet{rmode} we consider a variable $a$ which labels the (rotationally) deformed equipotential surfaces, and which is defined by the relation
\be
r=a[1+\epsilon(a,\theta)]
\ee
where $\epsilon$ is a function which represents the rotational deformation of the equilibrium structure from the spherical background model. To second order in the
slow-rotation approximation, the deformation can be cast in the form
\be
\epsilon=D_1(a)+D_2(a)P_2(\cos\theta)
\ee
where $P_2$ is the $l=2$ Legendre polynomial. For an $n=1$ polytrope we have \citep{Chandra}
\be
D_1=\frac{2}{\pi^2}\frac{M_0a\psi_1}{RM_r}\tilde{\omega}^2\;\;\;\;\mbox{and}\;\;\;\;
D_2=-\frac{1}{9}\frac{M_0a\psi_2}{RM_r}\tilde{\omega}^2\label{didue}
\ee
where $M_r$ is the mass contained within a radius $r$, and $R$ and $M_0$ are the radius and mass of the non-rotating star, respectively.
We have also defined
\be
\tilde{\omega}=\Omega\left(\frac{R^3}{GM_0}\right)^{1/2},\;\;\;\;\;\;\psi_1=1-\frac{\sin{y}}{y}\;,\;\;\;\;\;\;\psi_2=\frac{15}{y}\left[\left(\frac{3}{y^2}-1\right)\sin y-\frac{3}{y}\cos y\right]
\ee
where we have used the dimensionless variable  $y=a\pi/R$.
In terms of the variable $a$ the equations of hydrostatic equilibrium take the simple form
\be
\frac{1}{\rho(a)}\frac{dP(a)}{da}=-\frac{d\Phi_R(a)}{da}
\ee
with $\Phi_R=\Phi-\frac{1}{2}\Omega^2a^2\sin^2\theta$. Since the variable $a$ is associated with equipotential surfaces of $\Phi_R$, it follows that
the density profile for an $n=1$ polytrope has the same functional form as in the spherical case. That is, we have
\be
\rho=\rho_c\frac{\sin{y}}{y}\;\;\;\;\mbox{with}\;\;\;\;\rho_c=\frac{\pi M_0}{4R^3}
\ee
and we find that the mass of the rotating star is given by
\be
M=M_0\left[1+\frac{2}{\pi^2}{\tilde{\omega}}^2\left(\frac{\pi^2}{3}-1\right)\right]
\ee
Using this relation, we can impose that $M$ remain constant for all rotation rates and thus determine the central density as a function of $\tilde{\omega}$.
This then determines the value of the coordinate $a$ of the transition density where hyperons first appear. Figure \ref{radius} shows an example of the extent of the exotic core, for different rotation rates, for this analytic polytropic model.

\section{Estimating the hyperon bulk viscosity}

The aim of the present work is to estimate the effect
that hyperon bulk viscosity has on the unstable r-modes. In the core of a mature neutron star,
where we may expect a sizable hyperon population, several components are likely to be superfluid at temperatures below $\sim 10^9$ K.
This complicates the picture considerably as it becomes necessary to use a multifluid description of the system.
This, in turn, gives rise to, potentially, quite a large number of dissipation coefficients \citep{CQG, prep}
All calculations to date have considered single fluid systems,  including the effects of superfluidity only in the calculation of the reaction rates.
This is somewhat restricted as it means that the only dissipation coefficients that have been considered are the ``standard'' ones for bulk and shear
viscosity. A notable exception is the calculation of the ``superfluid'' bulk viscosity coefficients for a neutron star with a hyperon core by \citet{Gusakov2008},
and the application of these results to the damping of sound waves \citep{Gusakovsound}. These studies show that the additional damping coefficients can play a significant role.
In the following we shall estimate the superfluid bulk viscosity coefficients in the case of a core comprising neutrons, protons  and $\Sigma^{-}$ hyperons,
a system that turns out to closely resemble superfluid Helium \citep{helium}.

\subsection{Multifluid equations of motion}

Let us briefly outline the multifluid equations of motion for a system formed of neutrons ($\n$) and protons ($\p$) locked to the $\Sigma^{-}$ hyperons.
For simplicity, we neglect the presence of the neutral $\Lambda$ hyperons (which, if they are superfluid, would add another degree of freedom)
and only consider the bulk viscosity due to the process
\be
\n+\n \rightleftharpoons \p+\s^{-} \label{react}
\ee
This is, of course, not the only contribution to the hyperon bulk viscosity. However,  reactions involving $\Lambda$ hyperons
have been considered by  \citet{LO} and \citet{Gusakov2008}. Their results demonstrate that the inclusion of $\Lambda$ does not impact strongly on the
qualitative features of the bulk viscosity. Our main reason for ignoring the presence of $\Lambda$ hyperons is that we then have to consider only two fluids, the superfluid neutrons and a charge-neutral
conglomerate of protons and $\Sigma^{-}$ hyperons. In this case we can
use the analytic results of \citet{haensel} for the  bulk viscosity coefficients.

We shall assume that protons and $\Sigma^{-}$ hyperons are locked together by the Coulomb interaction, which proceeds on a much faster
timescale than the dynamical timescale we are interested in (i.e. the r-mode frequency).  We therefore consider one  single charge-neutral conglomerate. A full derivation and in detail discussion of the relevant equations of motion can be found in \citet{prep} and \citet{CQG}.
We will also ignore leptonic reactions, such as modified and direct URCA, as they proceed on a much longer timescale compared to the
 reaction in (\ref{react}) \citep{haensel}. Finally, we will (more or less) neglect the presence of electrons. They could easily be included  by assuming that they are also locked to the protons
(as in the usual model for the outer core of a neutron star \citep{Ma,Mb}), and that their inertia can be neglected \citep{MHD}. In the presence of
$\Sigma^{-}$ the number density of electrons is also depleted (due to overall charge neutrality), which means that they become less relevant in the deep core anyway.

Before locking the hyperons to the protons we have three distinct components.
Following \citet{CQG} we can write the momentum for each of  component as
\bear
\pi^\n_{i}&=&g_{ij}\left(m_\n n_\n v_\n^j-2[\alpha^{\n\p}w_{\n\p}^j+\alpha^{\n\s}w_{\n\s}^j]\right)\label{m11}\\
\pi^\p_{i}&=&g_{ij}\left(m_\p n_\p v_\p^j+2[(\alpha^{\n\p}+\alpha^{\p\s})w_{\n\p}^j-\alpha^{\p\s}w_{\n\s}^j]\right) \label{momp}\\
\pi^\s_{i}&=&g_{ij}\left(m_\s n_\s v_\s^j+2[(\alpha^{\n\s}+\alpha^{\p\s})w_{\n\s}^j-\alpha^{\p\s}w_{\n\p}^j]\right)\label{mom}
\eear
In these expressions,  $n_\X$ ($\X=\n,\p,\s$) is the number density of each species,  $v_\X^i$ is the velocity of each component and $w_{\X\Y}^i=v^i_{\X}-v^i_{\Y}$ represents
relative flows. The coefficients $\alpha^{\X\Y}$ describe the so-called ``entrainment'' effect, which leads to the momentum of a given species not being aligned with the individual velocity.
The equations of motion then take the form of coupled Euler equations (omitting for the moment the  effects of gravity);
\be
f_i^\X=\partial_t\pi^\X_i+\nabla_j\left(v_\X^j\pi^\X_i+D^{\X j}_i\right)+n_\X\nabla_i\left(\mu_\X-\frac{1}{2}m_\X v_\X^2\right)+\pi^\X_j\nabla_i v_\X^j
\ee
where $\mu_\X$ is the chemical potential of each species, $f_i^{\X}$ represents the sum of all external forces acting on the component $\X$ and $D^{\X j}_i$ represents the
dissipative part of the stress tensor, which includes shear and bulk viscosity \citep{CQG} .
In the following we  focus on hyperon bulk viscosity and thus only explicitly include the corresponding terms in the equations of motion.
These equations are complemented by continuity equations for each component, in the form:
\be
\partial_t n_\X+\nabla_i \left(n_\X v_\X^i\right)=\Gamma_\X
\ee
where $\Gamma_\X$ is the particle creation rate per unit volume for component $\X$. In general, $\Gamma_\X$ depends on all reactions involving the $\X$ particle species.
This will render the complete expressions quite complicated. This is another reason why we focus on the single reaction \eqref{react}.
Finally we assume overall baryon number conservation, i.e.
\be
\sum_{\X}\Gamma_\X=0
\ee
\subsection{The linearised problem}

Since we are interested in the r-modes we will follow \citet{rmode} and consider perturbations of a rotating background
in which all the fluids flow together. Assuming that the fluids are in chemical equilibrium we then have $\Gamma_\X=0$ in the background.
In writing the perturbation equations we shall assume, as discussed previously, that the  $\Sigma^{-}$ hyperons are also locked to the charged
component by the Coulomb interaction. In this case one has (indicating Eulerian perturbations with $\delta$ and Lagrangian perturbations with $\Delta$) $\delta w_{\n\s}^i=\delta w_{\n\p}^i$ and the problem is reduced to that of a two-fluid flow.
The equations of motion can be written in terms of two independent velocities which, following \citet{rmode}
we take to be  $\delta w_{\n\p}^i$ and the ``total'' velocity $\delta v^i$, defined as
\be
\rho \delta v^i=\sum_\X\rho_\X \delta v_\X^i
\ee
where $\rho_\X=m_\X n_\X$ is the mass density of each component and $\rho=\sum_\X\rho_\X$. We  also introduce the total pressure $p$ such that
\be
\nabla_i p = \sum_\X n_\X \nabla_i \mu_\X\label{pres}
\ee
In terms of these variables the perturbed Euler equations (including the dissipative terms due to bulk viscosity), to linear order and in a frame rotating with the star,
can be cast in the form:
\be
\partial_t (\rho\delta v_i)+\nabla_i\delta p+2\rho\epsilon_{ijk}\Omega^j\delta v^k-\frac{\delta\rho}{\rho}\nabla_i p=-\nabla^j \left(\sum_\X D^\X_{ij}\right)
\ee
\be
\partial_t\left[(1-\bar{\varepsilon}-\bar{\varepsilon}_2)\delta w^{\n\p}_i\right]+\frac{1}{2}\nabla_i\delta\tilde{\beta}+2\epsilon_{ijk}\Omega^j\delta w_{\n\p}^k=-\nabla^j\left( \frac{D_{ij}^\n}{\rho_\n}-\frac{D_{ij}^\s}{2\rho_\s}-\frac{D_{ij}^\p}{2\rho_\p}\right)
\label{Euler1}
\ee
and the continuity equations take the form
\be
\partial_t\delta n_b=-\nabla_j \left( n_b \delta v^j \right) -\frac{\Delta m}{m_\n}\nabla_i j^i
\label{continuo1}\ee
\be
\partial_t\delta x_\s ={ 1 \over n_b} \left(1+x_\s\frac{\Delta m}{m_\n}\right)\nabla_i j^i -\delta v^i\nabla_i x_\s+\frac{\Gamma_\s}{n_b}\\
\label{continuo2}
\ee
where we have defined
\bear
 j^i &=& n_b x_\s(1-y_\c) \delta w_{\n\p}^i \ , \\
 \Delta m &=& m_\s-m_\n \ , \\
 \bar{\varepsilon} &=& \frac{2(\alpha^{\n\p}+\alpha^{\n\s})}{\rho y_\c(1-y_\c)} \label{epsdef}\ ,\\
 \bar{\varepsilon}_2 &=& \frac{\Delta m}{m_\n}\frac{\alpha^{\n\p}-\alpha^{\n\s}}{\rho y_\c} \ , \\
 \tilde{\beta}&=&2\mu_\n/m_\n-\mu_\s/m_\s-\mu_\p/m_\p \ .
 \eear
We have also introduced the baryon number density $n_b$, the fraction $x_\X=n_\X/n_b$ (with $x_\c=(n_\s+n_\p)/n_b$) and the mass fraction $y_\c=(\rho_\s+\rho_\p)/\rho$. Note that $\tilde{\beta}$ does not vanish in the background. Rather, it is of order $\mathcal{O}(\Delta m/m_\n)$, as chemical equilibrium with respect to the reaction in (\ref{react}) implies $2\mu_\n-\mu_\s-\mu_\p=0$.
If we consider bulk viscosity to be the only dissipative process at work, we can write the dissipative contributions to the stress tensor as
 \beq
 D_{ij}=\sum_\X D^\X_{ij}&=&- g_{ij} \left[\zeta \nabla_l \delta v^l +{\zeta}^{\n\s} \nabla_l j^l\right]\label{stresstensor}\\
 \frac{D_{ij}^\n}{\rho_\n}-\frac{D_{ij}^\s}{2\rho_\s}-\frac{D_{ij}^\p}{2\rho_\p}&=&- g_{ij} \left[\tilde{\zeta}^{\n\s} \nabla_l \delta v^l +\tilde{\zeta}^\s \nabla_l j^l\right]
 \eeq
 This allows us to cast the Euler equations in the form
 \bear
&&\partial_t \left( \rho\delta v_i \right) +\nabla_i\delta p+2\rho\epsilon_{ijk}\Omega^j\delta v^k-\frac{\delta\rho}{\rho}\nabla_i p=\nabla_i\left[\zeta \nabla_l \delta v^l +{\zeta}^{\n\s} \nabla_l j^l)\right]
\label{Euler2a}\\
&&\partial_t\left[(1-\bar{\varepsilon}-\bar{\varepsilon}_2)\delta w^{\n\p}_i\right]+\frac{1}{2}\nabla_i\delta\tilde{\beta}+2\epsilon_{ijk}\Omega^j\delta w_{\n\p}^k=\nabla_i\left[\tilde{\zeta}^{\n\s} \nabla_l\delta v^l +\tilde{\zeta}^\s \nabla_l j^l\right]
\label{Euler2b}
\eear
As we can see, we now have four bulk viscosity coefficients, of which only three are independent (as expected from the analysis in \citet{prep}).
That is, we need to determine two bulk viscosity coeffients that are not present in the single fluid problem (only
the $\zeta$-term is present in the Navier-Stokes equations).
These extra coefficients can be calculated from the equation of state and the reaction rate $\Gamma_\s$.

Finally, we impose charge neutrality. As we are neglecting the electrons in the core, we have for the background
\be
x_\s\approx x_\p\approx \frac{x_\c}{2}
\ee
Meanwhile, for the perturbations charge neutrality leads to the condition
\be
\Delta x_\p=\Delta x_\s=\frac{\Delta x_\c}{2}
\label{charge}\ee
where $\Delta$ represents a Lagrangian perturbation. It is defined by
\be
\Delta=\delta+\mathcal{L}_\xi
\ee
where $\mathcal{L}_\xi$ is the Lie derivative with respect to the Lagrangian displacement $\xi^i$ associated with the co-moving motion, such that $\partial_t \xi^i=\Delta v^i$ (note that as we are in a two-fluid system it would also be possible to define a Lagrangian displacement associated with the counter-moving motion).

Solving the full set of equations (\ref{continuo1})--(\ref{Euler2b}), including the dissipative terms, is  still
a prohibitive task. We  thus follow the common strategy of assuming that the dissipative terms only
introduce a small deviation from the solution of the non-disipative problem, and  use this solution to estimate
the bulk viscosity damping timescale. The non-dissipative equations follow if we set $\Gamma_\s=\zeta=\zeta^{\n\s}=\tilde{\zeta}^{\n\s} =\tilde{\zeta}^\s=0$.
We also note that, since $\Gamma_\X=0$, we can rewrite \eqref{continuo2} as
\be
\partial_t\delta y_\s = { 1 \over \rho} \left(1+x_\s\frac{\Delta m}{m_\n} \right)\nabla_j \left\{ \rho y_\c[(1-y_\c)\delta w^j_{\n\p}]\right\} -\delta v^j\nabla_j y_\s \ .
\label{continuond}
\ee
These are, in fact, almost exactly the equations that were considered by \citet{rmode}. The only difference is
the entrainment dependence in equation (\ref{Euler2b}), where one has the extra term $\bar{\varepsilon}_2$ due to the difference in mass between neutrons and $\Sigma^{-}$ hyperons. The similarity
of the final equations obviously means that it is straightforward to adapt the method from \citet{rmode} to the present problem.

\section{Single-fluid bulk viscosity revisited}

Given our aim, we take as
starting point the study of
\citet{haensel}. Their results are particularly useful because they
provide relatively simple parameterised expressions for the
$\Sigma^-$ hyperon bulk viscosity coefficient also in the case of superfluid constituents.
Of course, these explicit expressions come at a price. They are based
on the bare-particle assumption, e.g. do not
consider the effective masses (of neutrons, protons and
hyperons). As emphasised by Haensel et al, this approximation is likely to be rather severe
and one may find that ``dressed particle'' effects affect the
results significantly. In order to allow for this possibility,
Haensel et al opted to leave a free parameter ($\chi$) in their formulae.
In our study, we will use the freedom associated with this parameter
to discuss the plausible range for the hyperon bulk viscosity.
This ``phenomenological'' approach to the problem is reasonable given the many uncertainties
associated with the supranuclear equation of state.  Finally, let us remark that the calculation of the bulk viscosity coefficient
is made assuming that the fluids are locked together. As discussed in section 5.1, this is equivalent to neglecting the dependance
of the reaction rate on the divergence of the relative velocities. Although this assumption is not physically justified,  we  make it for the sake of simplicity
 and in order to make progress on the r-mode problem.

To calculate the bulk viscosity coefficient due to the reaction $\n+\n\rightleftharpoons \p+\Sigma^{-}$ we  follow the procedure of \citep{Sawyer1989}.
In the single-fluid case the reaction rate $\Gamma_\s$ depends only on  the lag in instantaneous chemical potentials
\be
\Delta \beta=(2\Delta \mu_\n- \Delta \mu_\p- \Delta\mu_\s)
\ee
Then
\be
\Gamma_\s=-\lambda \Delta\beta \label{lambda}
\ee
where the coefficient $\lambda$ can be obtained from \cite{haensel}.
If we assume that the fluid is in equilibrium with respect to all other reactions, and impose that the perturbations maintain charge neutrality, i.e.
\be
\Delta n_\s=\Delta n_\p,
\label{charge2}
\ee
we can write the lag in chemical potentials as a function of two parameters, e.g. the total baryon number and the hyperon fraction. Thus,  $\Delta \beta= \Delta \beta(n_b, x_\s)$, from which we obtain
\be
\Delta
\beta=B\Delta x_\s+C\frac{\Delta n_b}{n_b}
\label{beta}
\ee
with
\be
B=\frac{\partial\beta}{\partial x_\s}\;\;\;\;\mbox{and}\;\;\;\;C=\frac{\partial\beta}{\partial n_b}n_b
\ee

By inserting equation (\ref{beta}) into the continuity equations (\ref{continuo1})--(\ref{continuo2}) and making use of the fact that $\Delta x_\c=2\Delta x_\s$, which follows from equation (\ref{charge}), we have
\beq
\partial_t\Delta n_b&=&-n_b\nabla_i \left( \Delta v^i \right)\\
\partial_t \Delta x_\s&=&-\lambda\frac{1}{n_b}\left(B\Delta x_\s+C\frac{\Delta n_b}{n_b}\right)\\
\eeq
If we assume a harmonic time dependance with frequency $\omega$ for the perturbations, these relations allow us to compute the hyperon fraction.
Finally, as the dissipation is given by the part of the pressure due to deviations from chemical equilibrium, we have (denoting with $\p_{eq}$ the pressure in chemical equilibrium):
(note that the variation is taken at constant baryon number density):
\be
p=p_{eq}+\frac{\partial p}{\partial x_\s}\Delta x_\s
\ee
which, inserted into the Euler equations (\ref{Euler2a}) and (\ref{Euler2b}) leads to the result of \citet{haensel};
\begin{equation}
\zeta_\Sigma = { C^2 n_b^2 \over |\lambda | B^2 } { 1 \over 1 + a^2}
\label{zeta1}\end{equation}
where
\begin{equation}
a = { \omega n_b \over |\lambda | B}
\end{equation}
Here $|\lambda| \propto 1/T$ (in non-superfluid matter) is
related to the relaxation time-scale of the relevant  reactions
which drive matter towards equilibrium.
We can then use the estimate of \citet{haensel} :
\be
a\approx 6.09 \left(\frac{m_n}{m_n^*}\right)^2\frac{m_p}{m_p^*}\frac{m_\Sigma}{m_\Sigma^*}\frac{\omega_4}{T_9^2\chi}
\left(\frac{n_b}{1\ \mbox{fm}^{-3}}\right)\left(\frac{1\ \mbox{fm}^{-3}}{n_\Sigma}\right)^{1/3}\left(\frac{100\mbox{MeV}}{B}\right)
\label{aa} \ee
where $T_9=T/10^9$ K and $\omega_4=\omega/10^4$ s$^{-1}$. In the following, we shall take as canonical values $m_\X^*=0.7 m_\X$ and $\chi=0.1$, as in \citet{haensel}. We will also focus on a stellar model
represented by an $n=1$ polytrope. The reason for this simplification is that we can make direct use of the analytical r-mode results of \citet{rmode}.
In future studies it would, of course, be desirable to calculate the coefficients in (\ref{zeta1}) consistently from a realistic equation of state and carry out
the mode-analysis for models constructed using the same equation of state. However, in order for this level of modelling to be meaningful one also has to
account for general relativity, c.f. the discussion in Section 2. In the case of the r-mode instability this has been done, but not for truly realistic equations of state \citep{Ruoff01,Ruoff02,Lockitch1,Lockitch, Yoshida, Gar}.
Further work is clearly needed.

Before we proceed, let us note that the bulk viscosity
is weak whenever the parameter $a$ is either very small or very large. That is, when the relaxation timescale associated with the reaction (\ref{react}) is significantly shorter
or longer than the oscillation timescale. This is natural since bulk viscosity is essentially a resonance effect associated with the dynamics and the reactions.
As a result, at constant
baryon number density $n_b$ and oscillation frequency $\omega$,
there are two asymptotic regimes;
\begin{equation}
\zeta_\Sigma^\mathrm{high}  \approx 4.1 \times 10^{32} { 1 \over \chi}
x_\Sigma^{-1/3} \left( {n_b \over 1 \mbox{ fm}^{-3}} \right)^{5/3}
    \left( {10^9 \mbox{ K} \over T} \right)^{2}\left(\frac{100\mbox{MeV}}{B}\right)^2\left(\frac{C}{100\mbox{MeV}}\right)^2 \ \mbox{ g/cm s}
\end{equation}
at high temperatures, and
\begin{equation}
\zeta_\Sigma^\mathrm{low}  \approx 6.3 \times 10^{29}  \chi
x_\Sigma^{1/3} \left( {n_b \over 1 \mbox{ fm}^{-3}}  \right)^{1/3}
    \left( { T \over 10^9 \mbox{ K} } \right)^{2} \left(
{10^4 \mbox{ s}^{-1} \over \omega  }
\right)^2\left(\frac{C}{100\mbox{MeV}}\right)
\ \mbox{ g/cm s}
\label{approx}\end{equation}
at low temperatures. Note that we could equivalently
refer to the asymptotic expressions as low- and high-frequency
approximations (at fixed temperature).
We expect these asymptotic expressions to be  good at low temperatures (below $10^9$ K) and  high temperatures (above $10^{10}$ K), but they are not valid in a temperature range where
\begin{equation}
\zeta_\Sigma^\mathrm{low} \approx \zeta_\Sigma^\mathrm{high}
\longrightarrow T_\mathrm{eq} \approx 5\times 10^9 \chi^{-1/2} x_\Sigma^{-1/6}
 \left( {n_b \over 1 \mbox{ fm}^{-3}}  \right)^{1/3} \left(
{ \omega  \over  10^4 \mbox{ s}^{-1}  }
\right)^{1/2}
\end{equation}
In this region the relaxation timescale is comparable to the
oscillation timescale, and one would not expect the
approximations above to be useful.

\section{Superfluidity}\label{super}

\begin{figure}
\centering
\includegraphics[height=8cm,clip]{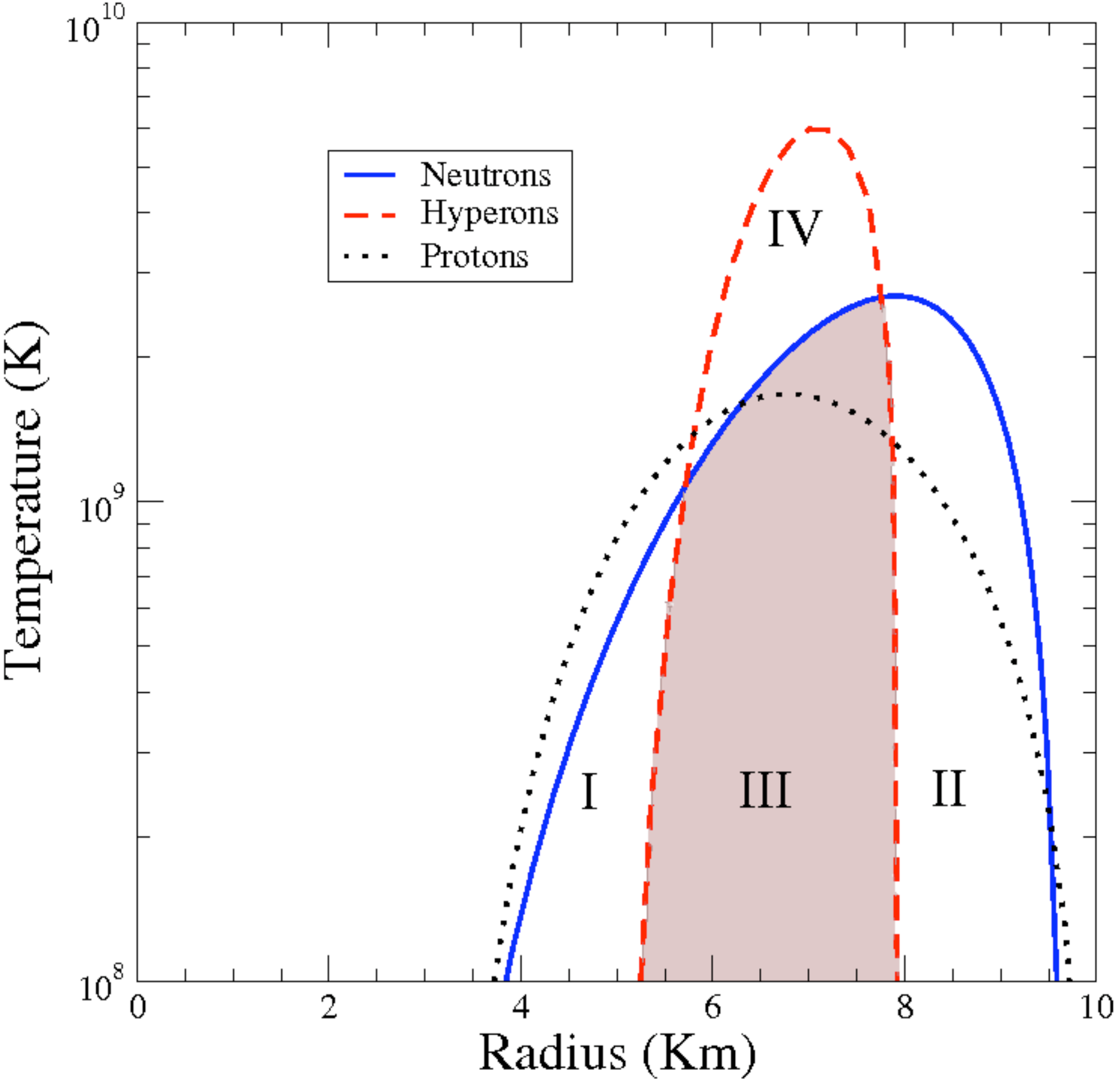}
\caption{Critical temperature for superfluidity of neutrons, protons and $\Sigma^{-}$ hyperons for a star with M=1.4 $M_{\odot}$ and R=$10$ km. For the hyperons we use the model of \citet{BB} while for the neutrons we use model "e"  and for the protons model "h" of \citet{nuclphys}. In regions I and II the neutrons are superfluid, but not the hyperons, and the system should exhibit multifluid behaviour. Region III, in which all components are superfluid, will also exhibit multifluid behaviour. In region IV only the hyperons are superfluid and all components should flow together. }
\label{gaps}
\end{figure}

At temperatures below $\sim 10^{10}$ K hyperons (and also neutrons and protons) are expected to form Cooper pairs and become superfluid. As we have already discussed, this can have a significant impact on the reaction rates and the bulk viscosity. The critical temperature below which each species is superfluid depends strongly on the density. In figure \ref{gaps} we show the critical temperature for neutrons (using model "e" from \citet{nuclphys}), protons (model "h" from \citet{nuclphys}) and $\Sigma^{-}$ hyperons, where we have assumed, as in \citet{Nayyar} that the pairing gap is the same as that for $\Lambda$ hyperons computed by  \citet{BB}.
This is essentially the gap deduced from BCS theory, so it does not account for medium effects etcetera.
This is worth keeping in mind since such effects are known to reduce the pairing gaps for neutrons and protons
significantly.
From the figure we see that there will be superfluid hyperons present below roughly $5\times10^9$~K.
We also learn that the transition will not take place until much lower temperatures are
reached in regions above 4-5 times nuclear density (above $n_b \approx 0.6$~fm$^{-3}$).
In fact, the figure indicates that the hyperons in the deep core may not be superfluid in
most astrophysical systems, e.g. the accreting neutron stars in LMXBs which are
expected to have core temperatures above $10^8$~K.
Note that  the pairing gaps we have used represent rough approximations to the detailed many-body results and different models lead to gap results that are  generally within
factors of a few of one another (for a review see \citet{nuclphys}). This will have an effect on the suppression of the reaction rates and the extent of the different regions described below, but
it is unlikely to affect the qualitative picture that emerges from our calculations.

When one or more of the species involved in the reactions that lead to the macroscopic bulk viscosity are superfluid, the reaction rates will be suppressed.
This can be accounted for by a multiplicative factor $|\lambda | \to
{\cal R} | \lambda|$. The form of this suppression factor depends strongly on which species involved in
reaction (\ref{react}) are superfluid and thus on the temperature and the region of the star one is considering.
In the region where only the hyperons are superfluid (region IV of figure \ref{gaps}) we can use the analytic reduction factor of \citet{haensel}, valid for singlet state pairing (note that we use the notation of \cite{haensel} and the symbols are not to be confused with those used in other sections of the paper):
\be
{\mathcal{R}_\s}=\frac{a^{5/4}+b^{1/2}}{2}\mbox{exp}(0.5068-\sqrt{0.5068^2+y^2})\label{hypsd}
\ee
where 
\be
y=\sqrt{1-\tau}\left(1.456-\frac{0.157}{\sqrt{\tau}}+\frac{1.764}{\tau}\right)\;\;\;\;\mbox{with}\;\;\;\;\tau=\frac{T}{T_c}
\ee
while $a=1+0.3118\;y^2$ and $b=1+2.556\;y^2$ and $T_c$ is the superfluid critical temperature for hyperons, which can be related to the superfluid gap by the relation $y=\Delta(T)/k_b T$, where $k_b$ is Boltzmann's constant.
Note that the above expression would be valid also in the case where only protons are superfluid and give us $\mathcal{R}_\p$, as long as one uses the corresponding critical temperature.
For the neutrons in the core, on the other hand, there will be triplet pairing, so to describe the reduction due to neutron superfluidity we shall consider the approximate factor of \citet{haensel}
\beq
{\mathcal{R}_\n}&=&(0.6192+\sqrt{0.3808^2+1.1561y^2})\times \mbox{exp}(0.7756-\sqrt{0.7756^2+y^2}) \nonumber\\
&&+0.18766\;y^2\times\mbox{exp}(1.7755-\sqrt{1.7755^2+4y^2})
\eeq
To describe the reduction rate $\mathcal{R}_{\n\p}$ needed in regions I and II of figure \ref{gaps} we would  need to account for  the reduction due
to the superfluidity of both neutrons and protons. \citet{haensel} do not provide an analytic expression for the reduction rate in these
regions, but \citet{Nayyar} found that the product of two single particle reduction factors (i.e. the factors
calculated for the case when only one species is superfluid) is a good approximation for the reduction rate in the case when both species are superfluid. In particular, they found that one could take  $\mathcal{R}_{\p\s}\approx\mathcal{R}_\p\mathcal{R}_\s$.
Assuming that this result is generally valid, we use
\be
\mathcal{R}_{\n\p}\approx \mathcal{R}_\n\mathcal{R}_\p
\ee
Note that, in the r-mode problem,
the prescription for the regions where only neutrons and protons are superfluid (regions I and II of figure \ref{gaps}) is not crucial,
as these regions do not dominate the contribution to the hyperon bulk viscosity damping. Region I is relatively unimportant because the r-mode eigenfunctions increase with radius, while the contribution
from region II is negligible as the hyperon number density falls steeply below $n_b \approx 0.32$~fm$^{-3}$ for our model EOS.
However, we need a prescription for the region where all the constituents are superfluid (region III of figure \ref{gaps}). In this region the reduction factor of \citet{haensel} can only be evaluated numerically, so we shall use the prescription
\be
\mathcal{R}_{\n\p\s}\approx \mathcal{R}_\n\mathcal{R}_\p\mathcal{R}_\s
\ee
which reproduces the qualitative features of the result of \citet{haensel}. In particular, the suppression becomes stronger when all the species
are superfluid.

\begin{figure}
\centering
\includegraphics[height=7cm,clip]{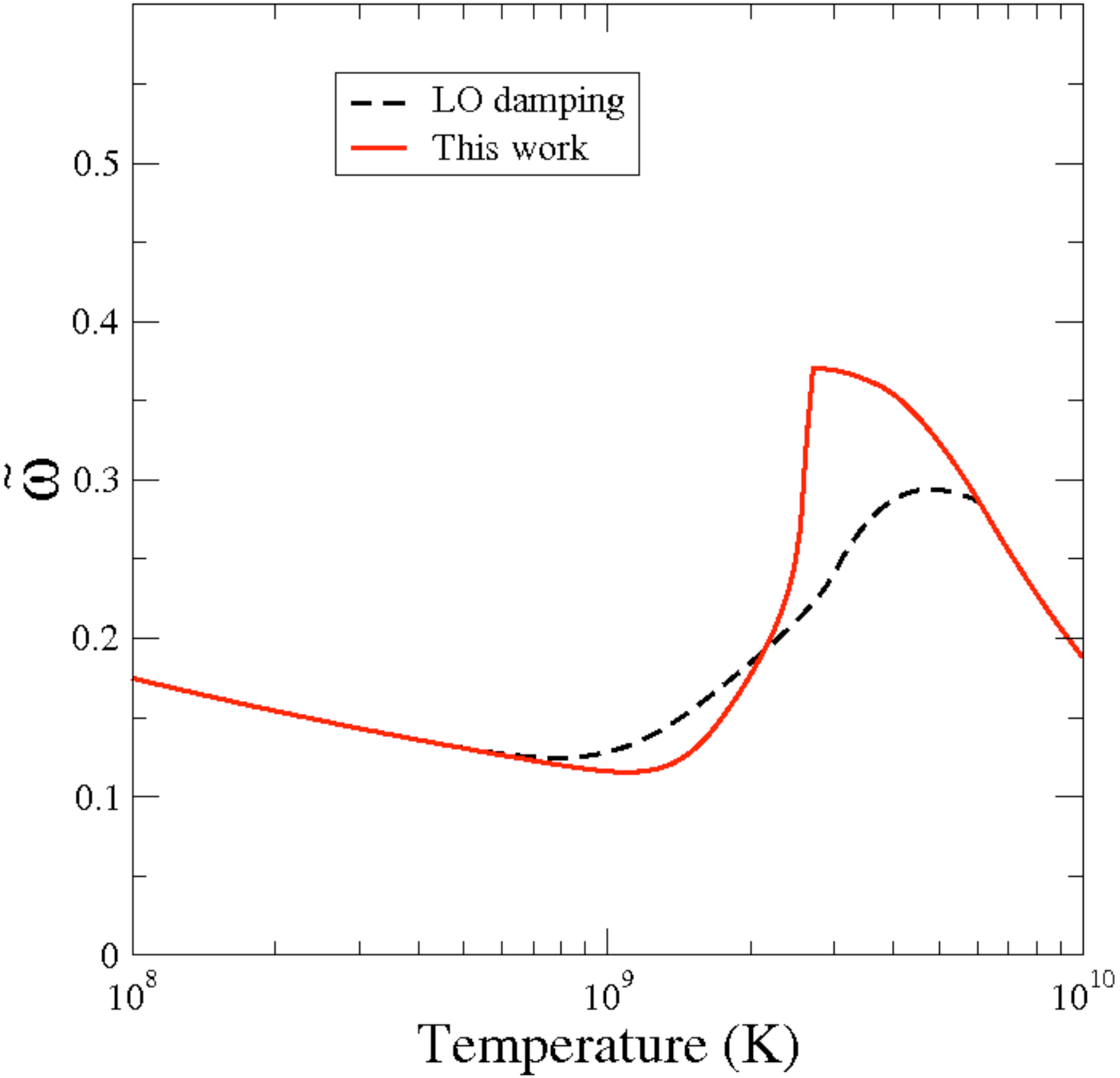}
\caption{The r-mode instability window in the single fluid case including the effect of hyperon bulk viscosity. We compare the superfluid reduction rate used in this work with the approximation of \citet{LO} (LO in the figure). It is clear that the approximation of \citet{LO} is reasonably accurate at lower temperatures when all species are superfluid. In the higher temperature region we use the exact reduction rate of \citet{haensel} for the case where only hyperons are superfluid, which leads to a smaller reduction factor and thus a slightly smaller instability window.
}
\label{damp}
\end{figure}

As a means of comparison we shall examine the effect of our prescription on the r-mode instability. To do this we shall calculate the standard single-fluid r-mode solution, as described by \citet{rmode}, and calculate the critical frequency at which hyperon bulk viscosity stops the mode from growing unstable (cf. section 6 for more details). In figure \ref{damp} we plot the critical r-mode frequency curve and compare the result obtained with our prescription for the superfluid suppression to that obtained with the approximate reduction rate used by \citet{LO}
\begin{equation}
{ 1 \over {\cal R}} =e^{\Delta/kT}
\label{dampLO}
\end{equation}
where $\Delta$ is the hyperon pairing gap. The comparison shows  that the approximation  (\ref{dampLO}) is  reasonably accurate, although it is also
clear that the more detailed reduction rate in (\ref{hypsd}) leads to a slightly smaller reduction factor.

 To complete our description we also need to specify in which region to use the one-fluid bulk viscosity timescale of equation (\ref{1fluid}) and in which to use the two-fluid timescale of equation (\ref{2fluid}). Technically, as the hyperons are locked to the protons, the system will exhibit two-fluid behaviour whenever the \underline{neutrons} are superfluid (regions I, II and III of figure \ref{gaps}). However, as we have mentioned above, we will only consider regions I and III, and neglect region II as it is essentially hyperon-free and does not contribute to the bulk viscosity. The whole star could of course be included and the picture be made consistent by using the full results for the superfluid reduction rates of \citet{haensel}, but such a calculation is beyond the scope of this qualitative study. 

\subsection{ Superfluid bulk viscosity}

To obtain a complete description of the bulk viscosity we need to have an estimate not only of the ``usual'' bulk viscosity coefficient, but also of the two extra ``superfluid'' ones. We will now
 use the standard bulk viscosity coefficient of \citet{haensel}, discussed in the previous section, to estimate the superfluid bulk viscosity
coefficients for the reaction $n+n\rightleftharpoons p+\Sigma^{-}$. This strategy obviously has some flaws, as the reaction rates and coefficients of \citet{haensel}
are calculated in the one fluid approximation, in which the reaction rate $\Gamma_\s$ depends only on  the instantaneous lag between the  chemical potentials.
For our multifluid system the reaction rate should depend, in general, also on the divergence of the relative velocity. That is, it should  take the form \citep{CQG}
\be
\Gamma_\s=-\lambda_w \Delta \beta-\tau_w\nabla_i \Delta w_{\n\s}^i
\ee
In the following we shall assume that $\lambda_w=\lambda$ (i.e. the same as in the one fluid case) and that $\tau_w=0$. In practice, we are assuming that the
variation of the reaction rate with respect to the single fluid result is small. This assumption is not necessarily justified but as there is, to the best
of our knowledge, presently no calculation of the coefficients $\lambda_w$ and $\tau_w$ it is the only option if we are to make progress.
For future applications it would be highly desirable to have a calculation of  the reaction rates for the full superfluid system and estimates of the impact of the parameter $\tau_w$ on the r-mode damping timescale.
Given the reaction rate $\Gamma_\s$,  we  follow the strategy of the previous section  and expand the pressure and difference in chemical potentials around a
 background solution at equilibrium. The continuity equations now have an extra term due to the difference in velocity between the neutrons and the charged component
\bear
\partial_t\Delta n_b&=&-n_b\nabla_i\Delta v^i-\frac{\Delta m}{m_n}\nabla_j\left[ n_b x_\s(1-y_\c) \Delta w^j_{\n\p}\right]\\
\partial_t \Delta x_\s&=&-\frac{\lambda}{n_b}\left(B\Delta x_\s +C\frac{\Delta n_b}{n_b}\right) + \frac{\nabla_j}{ n_b}\left[n_b x_\s(1-y_\c) \Delta w^j_{\n\p}\right]
\eear
Note that we have expressed the continuity equations in terms of Lagrangian perturbations, but as we are working in the rotating frame and have chosen a co-moving background, we have $\Delta w^i_{\n\p}=\delta w^i_{\n\p}$ and $\Delta v^i=\delta v^i$.
Following the analysis of the previous section we can expand the parts of the pressure and $\beta$ which depend on the deviations from chemical equilibrium;
\be
\delta p=\delta p_{eq}-\zeta \nabla_i \delta v^i -{\zeta}^{\n\s} \nabla_i j^i
\ee
and
\be
\delta \tilde{\beta}=\delta \tilde{\beta}_{eq}-\left(\frac{\partial\tilde{\beta}}{\partial x_\s}\right)_{n_b}\left(\frac{\partial p}{\partial x_\s}\right)^{-1}_{n_b}\left[{\zeta}\nabla_i \delta v^i +{\zeta}^{\n\s}\nabla_i j^i\right] \label{zetasup}
\ee
where we recall that $j^i=n_b x_\s(1-y_\c) \delta w_{\n\p}^i$, and $\zeta$ is the same coefficient as in equation (\ref{zeta1}).
Using equation (\ref{zetasup}) in the Euler equations (\ref{Euler2b}) we can infer the following relation
\be
\frac{\tilde{\zeta}^{\n\s}}{\tilde{\zeta}^\s}=\frac{\zeta}{\zeta^{\n\s}}
\label{tre}\ee
which, as expected from the Onsager symmetry principle \citep{prep}, shows that in fact only three of the coefficients are independent.
Moreover, we can rewrite equation (\ref{zetasup}) as
\be
\tilde{\beta}=\tilde{\beta}_{eq}-\left(\frac{\partial\tilde{\beta}}{\partial \beta}\right)_{p}\left[\bar{\zeta}^{\n\s} \nabla_i\delta v^i +{\zeta}^\s \nabla_i j^i+\bar{\zeta}^{\n\s}\frac{\Delta m}{m_\n n_b}\nabla_i j^i\right] \label{zetasu2}
\ee
which allows us to identify the coefficients:
\be
{\zeta}^{\n\s}=-\frac{\zeta}{n_b}\left[\frac{B}{C }\left(1+x_\s\frac{\Delta m}{m_n}\right)-\frac{\Delta m}{m_\n}\right]=\bar{\zeta}^{\n\s}\left(1+x_\s\frac{\Delta m}{m_n}\right)+\zeta\frac{\Delta m}{m_\n n_b}
\ee
\be
{\zeta}^\s=\zeta\frac{B^2}{C^2 n_b^2}\left(1+x_\s\frac{\Delta m}{m_n}\right) \label{zeds}
\ee

If we identify $\zeta$ with the coefficient obtained by \citet{haensel}, i.e. the one given in \eqref{zeta1}, the above relation allows us to
determine the other coefficients from the equation of state, and evaluate the bulk viscosity damping timescale.
Basically, this is equivalent to assuming that the reaction rates are the same as in the single fluid case.
The coefficients needed in equation (\ref{Euler2b}) are then:
\be
 \tilde{{\zeta}}^{\n\s}=\frac{1}{2}\left(\frac{\partial\tilde{\beta}}{\partial\beta}\right)_p{\bar{\zeta}}^{\n\s}
 \ee
 \be
 \tilde{{\zeta}}^{\s}=\frac{1}{2}\left(\frac{\partial\tilde{\beta}}{\partial\beta}\right)_p\left({\zeta}^{\s}+\frac{\Delta m\bar{\zeta}^{\n\s}}{m_\n n_b}\right)
 \ee
Finally, we can compare our results to those obtained from the relativistic formulation of \citet{Gusakov2008}. This is clearly a rough comparison,
as we have neglected both relativistic effects and the effect of $\Lambda$ hyperons but, if we for simplicity neglect entrainment and terms of the order of
$\Delta m/m_\n$, we can identify the dissipative stress tensor $D_{ij}$ in equation (\ref{stresstensor}) with the relativistic analogue in equation (63) of \citet{Gusakov2008}. Comparing the two expressions we find that the various
coefficients agree reasonably well.

\section{The r-mode instability window}\label{rmode}

The stability of an r-mode can be determined by estimating the gravitational-wave driving and viscous damping timescales.
The result is usually illustrated in terms of the critical frequency at which the driving and damping timescales are equal. In our case, the instability curve
is obtained by solving for the roots of
\be
\frac{1}{\tau_{gw}}+\frac{1}{\tau_{Hb}}+\frac{1}{\tau_{sv}}+\frac{1}{\tau_{Ek}}=0
\ee
where $\tau_{Hb}$ is the hyperon bulk viscosity damping timescale, $\tau_{sv}$ is the shear viscosity
damping timescale, $\tau_{Ek}$ is the damping timescale due to an Ekman layer at the base of the crust and $\tau_{gw}$ is the gravitational-wave growth timescale, which for an $n=1$ polytrope and the $l=m=2$ r-mode, is given by \citep{review}:
\be
\tau_{gw}=-47\left(\frac{M}{1.4M_{\odot}}\right)^{-1}\left(\frac{R}{10 \mbox{km}}\right)^{-4}\left(\frac{P}{1\mbox{ms}}\right)^{6}s
\ee
(the sign implies that the mode is growing).
The nature of the shear viscosity damping will depend on which region of the star we are considering, and one should also note that in a multifluid system there will, in general,
be more shear viscosity terms than in the single-fluid problem. 
The exact nature of the scattering processes that give rise to shear viscosity depends on which species are present and whether they are superfluid.
Shear viscosity in a hyperon core has not, to the best of our knowledge,  been studied in the literature.
Hence, we cannot at this stage consider hyperon scattering processes. 
We could, however, quantify the effect that the presence of hyperons has on the standard scattering processes.
In particular, we expect the electrons to be severely depleted in the presence of $\Sigma^-$, thus weakening the shear from electron-electron and proton-electron scattering. This should reduce the shear viscosity in the part of 
the star where neutrons, protons and $\Sigma^-$ are all superfluid. In regions where the neutron are normal, e.g. the deep core, we also know that neutron-neutron scattering will dominate and we can use the estimates of \citet{nuclphys}.
These details may not be that relevant, however, since the various scattering processes lead to weaker 
damping than the shear associated with the Ekman layer at the crust-core boundary \citep{BU,LOU,LU}. Since there will 
be no electron depletion in the outer core, we can use the standard result for the boundary layer damping. 
Thus we take
\be
\tau_\mathrm{Ek} \approx 3\times 10^5\left(\frac{T}{10^9 \mathrm{K}}\right)\left(\frac{10\ \mathrm{km}}{R}\right)^2\left(\frac{P}{1\ \mathrm{ms}}\right)^{1/2}\ \mathrm{s}
\ee
We arrive at this estimate by taking the simple constant density estimate of \citet{review} for a $M=1.4M_{\odot}$ neutron star, corrected for a ``slippage'' factor $\mathcal{S}_c$=0.05, as defined by \citet{eck2}. It has been shown by \citet{eck1} that one should expect the constant density estimate to only differ by factors of a few from the result for a stratified model. Hence, it should be a reasonable approximation. Having said that, it is absolutely clear that the boundary layer issue needs more detailed scrutiny. Based on our current understanding, 
the physics in the crust-core transition region dictates the r-mode damping at the temperatures that are 
relevant for mature neutron stars. Yet, our understanding of the effect that superfluidity and superconductivity 
may have on the boundary layer is far from complete \citep{Mend,TThesis}. These issues are
of central importance to neutron star dynamics, and more detailed studies should be encouraged.

\begin{figure}
\centering
\includegraphics[height=7cm,clip]{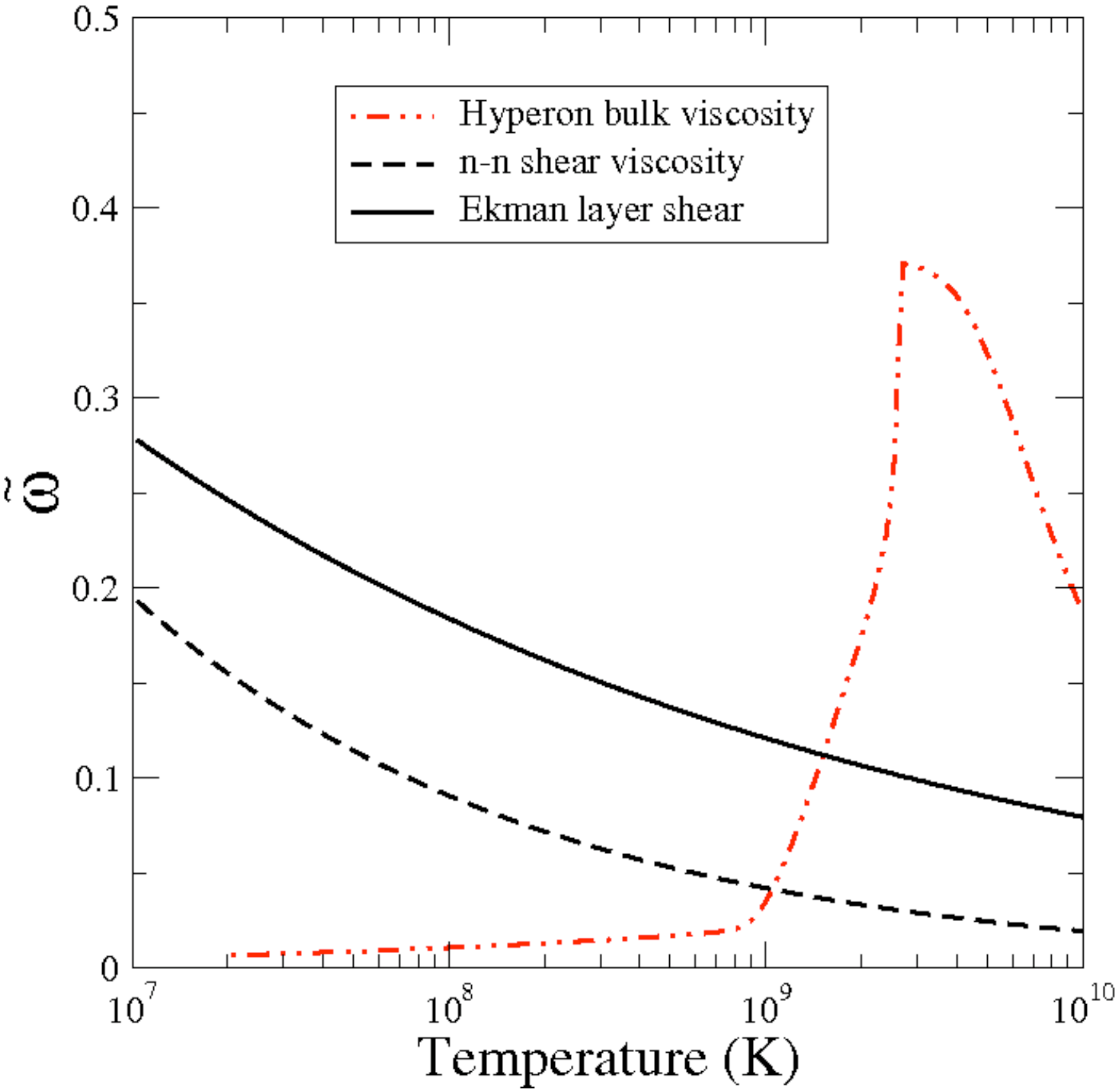}
\caption{We compare the critical curves for shear viscosity due to n-n scattering and for shear due to an Ekman layer, as described in the text. The damping from the Ekman layer is always stronger in the region of interest. For reference we have also included the critical timescale for hyperon bulk viscosity (with suppression of the reaction rates included, following the prescription described in the text).}
\label{shear}
\end{figure}

\begin{figure}
\centering
\includegraphics[height=6cm,clip]{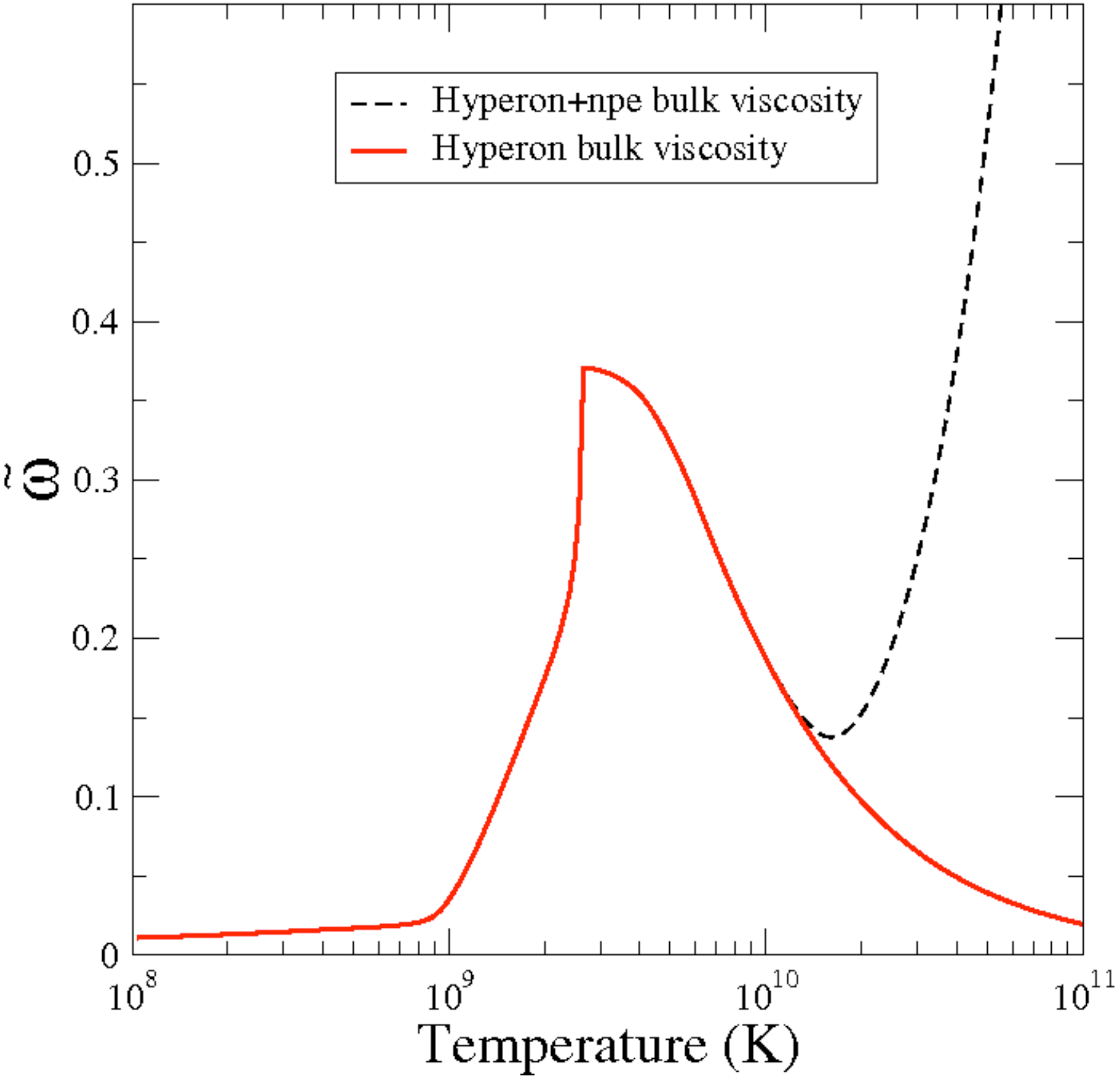}
\includegraphics[height=6cm,clip]{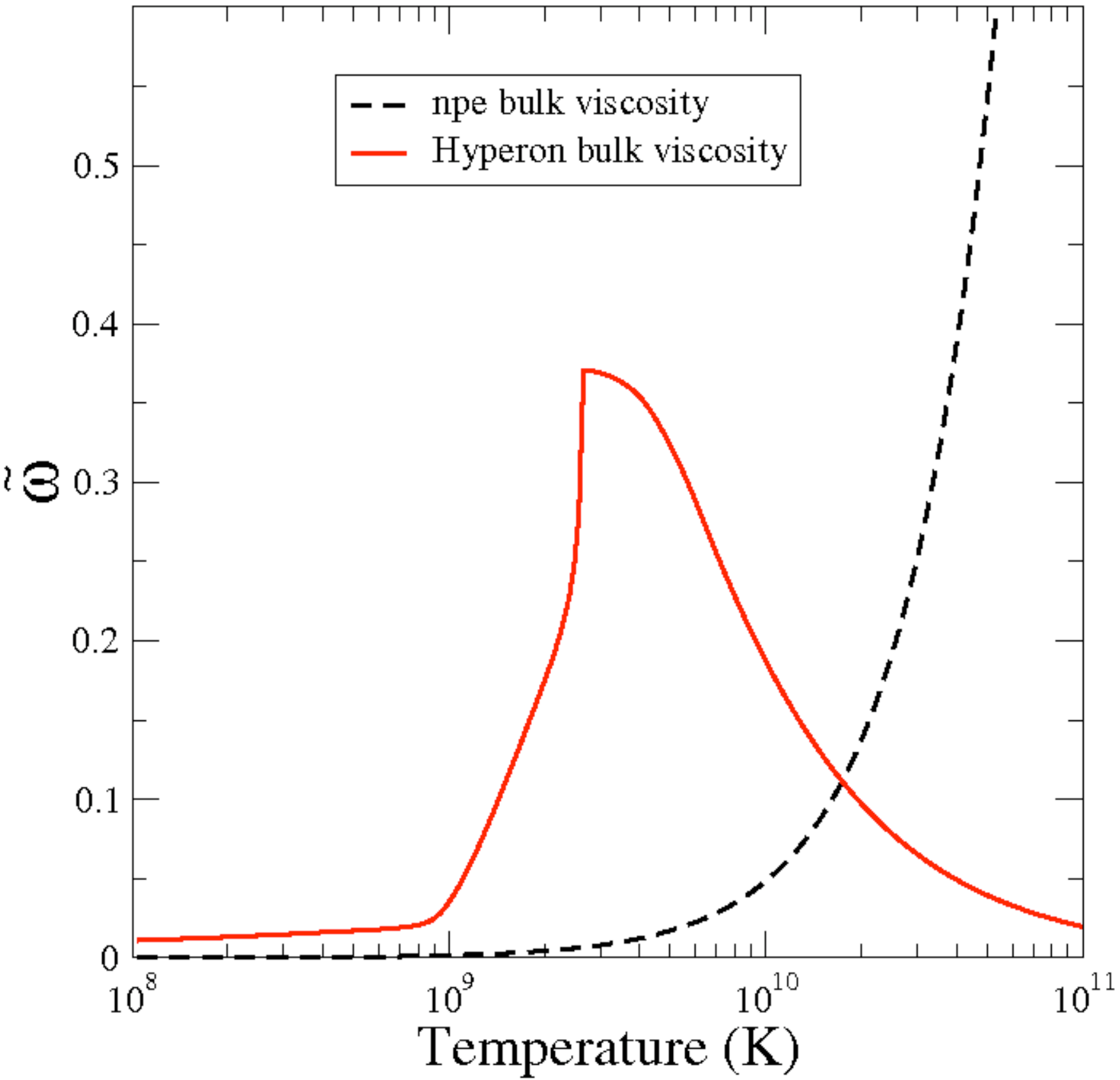}
\caption{In the left panel we show the effect on r-mode instability window of including npe bulk viscosity in the outer layer of the star. As is clear from the figure the effect is completely irrelevant below $\approx 10^{10}$ K, above which the contribution due to modified URCA reactions becomes dominant. This can easily be understood from the right hand side panel where we compare the damping timescale for hyperon bulk viscosity and for npe bulk viscosity. It is obvious from the figure that the npe bulk viscosity damping timescale is always much longer in the region of interest (below $\approx 10^{10}$ K).}
\label{rot1}
\end{figure}

The hyperon bulk viscosity damping timescale is given by
\be
\frac{1}{\tau_{Hb}}=-\frac{1}{2 E}\left(\frac{dE}{dt}\right)
\ee
To leading order in rotation, the energy of the mode takes the form \citep{rmode}
\be
E_k=\frac{1}{2}\int \rho\left[\delta v^2+(1-\bar{\varepsilon})y_\c(1-y_\c) \delta w_{\n\p}^2\right] dV
\ee
where we recall that  $y_\c=y_\p+y_\s$ and the definition \eqref{epsdef}.
In the one fluid region (which also includes the deep core), where the neutrons flow together with the protons and hyperons, we can write
\be
\frac{\partial E}{\partial t}=-\int \zeta(\nabla^i \delta v^i)^2dV
\label{1fluid}
\ee
where the bulk viscosity coefficient $\zeta$ is taken to be that of \citet{haensel} in the hyperon core, while in the outer layers of the neutron star we take the value given by \citet{Sawyer1989}, appropriate for modified URCA reactions in neutron, proton and electron (npe) matter
\be
\zeta=6\times 10^{25} \left(\frac{\rho}{10^{15}\mbox{g/cm$^3$}}\right)^2 \left(\frac{T}{10^9\mbox{K}}\right)^6 \omega_r^{-2}
\label{zetanormal}\ee
where $\omega_r$ is the r-mode frequency in the rotating frame. This leads to 
the r-mode damping timescale [where we have corrected an error in the numerical prefactor of the result from \citet{review}]
\be
\tau_{bv}\approx 8.6 \times 10^{11}\left(\frac{M}{1.4 M_{\odot}}\right)\left(\frac{R}{10\mbox{km}}\right)^{-1}\left(\frac{P}{1 \mbox{ms}}\right)^2\left(\frac{T}{10^9\mbox{K}}\right)^{-6} s\label{npe}
\ee
We note that although we have added the npe bulk viscosity contribution for completeness it is, in fact, irrelevant in the temperature range of interest for mature neutron stars. This is evident from the results in figure \ref{rot1}. The result is natural given the temperature scaling in \eqref{zetanormal} and the fact that the resonance associated with the modified URCA reactions is located at higher temperatures (above $10^{10}$~K) compared to the hyperon resonance (at a few times $10^9$~K) (note, however, that if a sizable proton fraction is present in the core, then direct URCA reactions can contribute to the bulk viscosity in the region where all baryons are superfluid and the hyperon reactions are reduced \citep{haensel}).

Let us now consider the two fluid region, where one has a host of extra terms due to the separate fluxes of the neutrons and the charged particles. In this region we
need to evaluate
\begin{multline}
\frac{\partial E}{\partial t}= -\int \zeta(\nabla^i \delta v^i)^2+ \frac{1}{2}{\zeta}^{\n\s}\left[\left({\nabla_i}\delta v^i\right)\left({\nabla^i}{ j}_i^{*}\right)+c.c.\right]+\\
+{f(\bar{\epsilon})}\left\{\tilde{\zeta}^\s\left[\left(\nabla_i j^i\right)\left(\nabla^i \tilde{j}_i^{*}\right)+c.c.\right]+\tilde{\zeta}^{\n\s}\left[\left(\nabla_i \tilde{j}^i\right)\left(\nabla^i \delta v_i^{*}\right)+c.c.\right]\right\} dV\\
=-\omega^2\int \zeta \left(\frac{\Delta p}{\Gamma p}\right)^2-\frac{{\zeta}^{\n\s}}{2(1+x_\s\frac{\Delta m}{m_\n})}\left[\left(\frac{\Delta\rho}{\rho}\right)(n_b\Delta x_\s)^{*}+c.c.\right]+\\
+{f(\bar{\varepsilon})}\left\{\frac{\tilde{\zeta}^\s}{(1+x_\s\frac{\Delta m}{m_\n})^2}\left[(\rho\Delta y_\s)(n_b\Delta x_\s)^{*}+c.c.\right]-\frac{\tilde{\zeta}^{\n\s}}{(1+x_\s\frac{\Delta m}{m_\n})}\left[\left(\rho\Delta y_\s\right)\left(\frac{\Delta \rho}{\rho}\right)^{*}+c.c.\right]\right\} dV \label{2fluid}
\end{multline}
where $c.c$ indicates the complex conjugate, $2 \tilde{j}^i=\rho y_\c(1-y_\c) \delta w_{\n\p}^i$ and $f(\bar{\varepsilon})=(1-\bar{\varepsilon})/(1-\bar{\varepsilon}-\Delta\bar{\varepsilon})$. Note that, following \citet{rmode} we have assumed that $\delta w^{\n\p}_i$ vanishes identically outside the two fluid region and that $\Delta p=\Delta x_\s=0$ at the surface of the star, which is equivalent to all the components having the same surface. This condition is ``natural'' as we do not expect the outer regions of the crust to be superfluid.
We have  taken $\bar{\varepsilon}$ to be a constant, although this is not necessarily, realistic, and in the following we will assume that, as an approximation to the results of \citet{Gusakov2009}, $\alpha^{\n\c}=\alpha^{\n\s}+\mathcal{O}(\Delta m/m_\n)$ and thus $f(\bar{\varepsilon})= 1+\mathcal{O}(\Delta m/m_\n)^2$. This is clearly an approximation, that follows from assuming (as described in appendix A) that the coefficients of the relativistic entrainment matrix are constant. Use of the full density entrainment parameters from \citet{Gusakov2009}, without neglecting higher order terms in  $\Delta m/m_\n$, would produce a weak entrainment dependence of the damping timescale.

\subsection{Model EOS}

In order to assess the superfluid hyperon bulk viscosity effect on the r-modes, we will make  use of the analytic solution by \citet{rmode}. This means that we
assume that the overall density profile is that of  an $n=1$ polytrope, in which case the solution for the comoving degree of freedom in the superfluid problem is completely independent
from the countermoving motion (which is, in fact, driven by the comoving part).
In order to evaluate the integral in (\ref{2fluid}) we need to consider the speed of sound, which for an $n=1$ polytrope of the form $p=K\rho^2$ is
\be
c_s^2=2K\rho
\ee
We shall also assume, as an ``approximation'' to case III EOS of \cite{Glen}, that in the core the hyperon number density follows a relation of the form
\be
x_\s=q_1 n_b+q_2
\ee
with $q_1=0.114$~fm$^3$ and $q_2=-0.036$. 
This allows us to consider a simple model that is linear in the number density of baryons, with the key feature that there are no hyperons below $n_b\approx 0.319 $ fm$^3$. A more realistic model would give a steep drop in the number density of hyperons close to the transition density, but as we are not using a realistic equation of state our linear approximation is sufficient.
With the use of the thermodynamical identity
\be
\frac{\partial{p}}{\partial x_\s}=-n_b^2\frac{\partial\beta}{\partial n_b}
\ee
one can then obtain
\beq
C&=&-\frac{c_s^2m_\n}{q_1 n_b}\left[1+\frac{\Delta m}{m_\n}\left(x_\s+n_b\frac{\partial x_\s}{\partial n_b}\right)\right]\\
B&=&-\frac{c_s^2m_\n}{q_1^2 n^2_b}\left[1+\frac{\Delta m}{m_\n}\left(x_\s+n_b\frac{\partial x_\s}{\partial n_b}\right)\right]
\eeq
where we have used the relation $\rho=m_\n n_b(1+x_\s\Delta m/m_\n)$.
We can now simplify the integral in (\ref{2fluid})  making use of:
\beq
\Delta x_\s&=&\frac{\partial x_\s}{\partial n_b}\Delta n_b
\eeq
Furthermore, we can write the mass fraction $y_\s$ as
\be
y_\s=\frac{m_\s}{m_\n}x_\s\left(1-\kappa\right)\;\;\;\;\mbox{and}\;\;\;\;\rho=m_\n n_b(1+\kappa)
\label{kappa}
\ee
where we  expand in terms of the parameter $\kappa=\Delta m/{m_\n}x_\s$. This is justified as $\kappa\approx 0.03$ for the densities we shall consider. From (\ref{kappa}) we can then conclude that
\be
\Delta y_\s=\frac{m_\s}{m_\n}\Delta x_\s \left(1-2\kappa\right)
\ee
Finally, charge neutrality (\ref{charge}) implies $\mu_\p\approx\mu_\s$ and from (\ref{pres})
\be
\left(\frac{\partial\mu_\n}{\partial{\beta}}\right)_p=x_\s
\ee
which leads to
\be
\left(\frac{\partial\tilde{\beta}}{\partial\beta}\right)_p=\frac{1}{m_\n}\left[1-\frac{m_\n}{m_\s}{\kappa}\frac{(1-x_\c)}{x_\c}\right]
\ee
and
\be
\frac{\Delta n_b}{n_b}=\frac{\Delta\rho}{\rho}\left[ 1-\kappa\frac{\partial x_\s}{\partial n_b}\frac{n_b}{x_\s} \right]
\ee
Using the values of the bulk viscosity coefficients from equation (\ref{zeds}) we can now write:
\be
\frac{\partial E}{\partial t}=-\omega^2\int_{\mathcal{V}_2} \left(\frac{\Delta p}{\Gamma p}\right)\zeta_\mathrm{eff} dV\label{2fluidbulk}
\ee
with
\be
\zeta_\mathrm{eff}\approx4\left[1+\frac{\Delta m}{m_\n}\frac{(1-x_\c-6\frac{\partial x_\s}{\partial n_b}n_b)}{4}\right]\zeta=4\left[1+\frac{\Delta m}{m_\n}\frac{(1-x_\c-6\alpha n_b)}{4}\right]\zeta
\ee

The integral in (\ref{2fluidbulk}) is considerably simpler than the original expression in (\ref{2fluid}). In particular, it only involves calculating the integral of the quantity $\frac{\Delta P}{\Gamma P}$ for the r-mode.
As discussed above, and presented in detail in \citet{rmode}, the co-moving degrees of freedom decouple form the counter-moving ones to second order in rotation, so only a solution to the single fluid r-mode problem is needed. In particular, as we are considering an $n=1$ polytrope we shall use the explicit analytic solution in section 4.6 of \cite{rmode}.
Note that the co-moving r-mode solution sources the counter-moving motion, thus giving rise to a relative flow and the additional bulk viscosity coefficients. 

\section{Results}

\begin{figure}
\centerline{\includegraphics[height=7cm,clip]{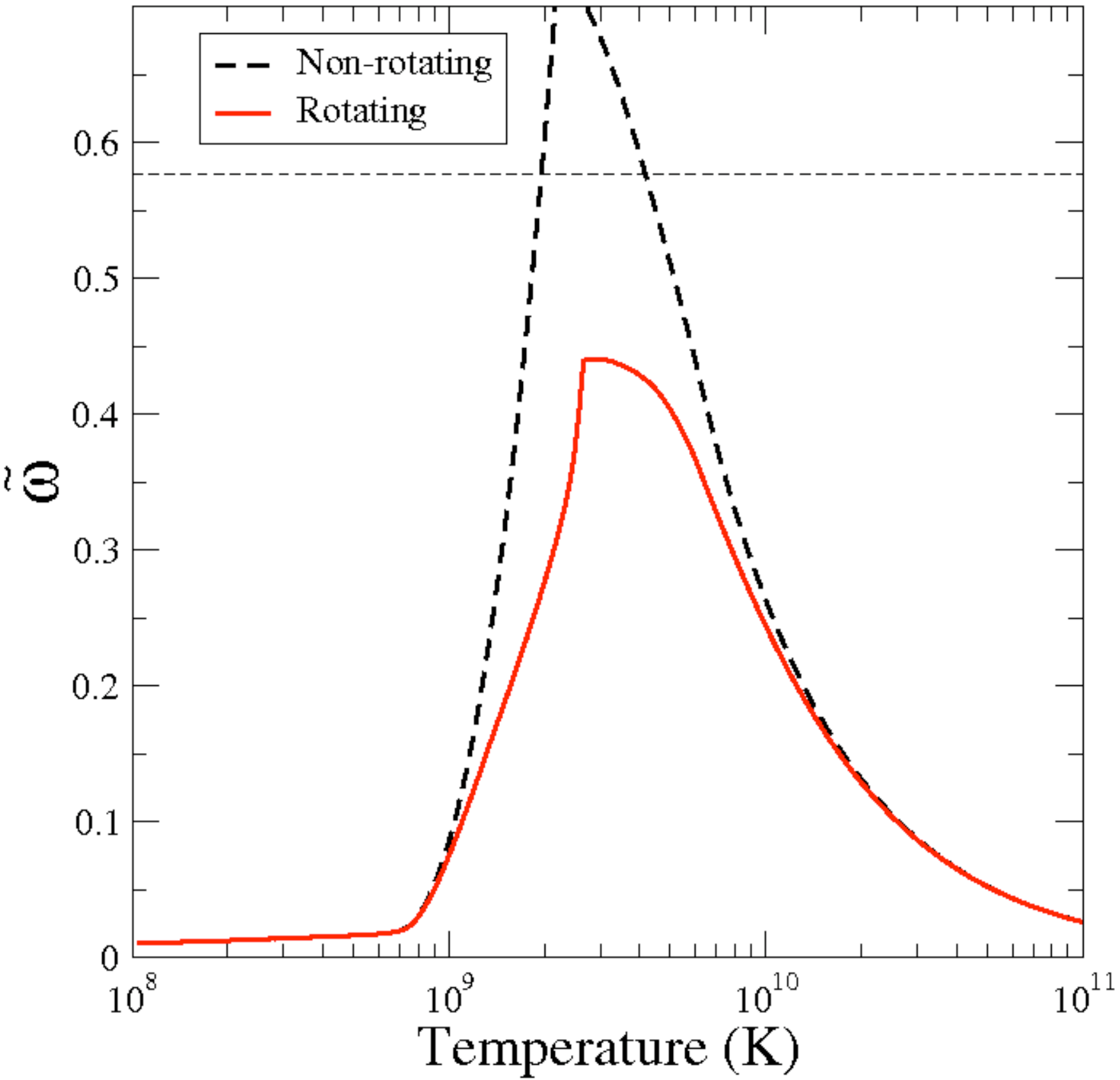}}
\caption{The critical curve due to hyperon bulk viscosity in the single fluid case, for a star with $M=1.4 M_{\odot}$ and $R=12.5$ km with a comparison between the non-rotating case and the case with the rotational corrections included. The straight horizontal line represents the Keplerian breakup velocity.}
\label{unstable}
\end{figure}

Before moving on to estimate the effect of the superfluid bulk viscosity, let us remark on the effect that rotation has on the instability window. In figure \ref{unstable} we show an example of the r-mode instability window, in the one fluid limit, for a stellar model with $M=1.4 M_{\odot}$ and $R=12.5$~km, using the hyperon bulk viscosity coefficient of \citet{haensel}. We compare the critical frequency with rotational corrections included in the size of the core and the frequency obtained without including them. It is immediately obvious that there is a significant difference for large rotation rates; in particular there is a range of temperature where the instability would be completely suppressed if we considered the non-rotating result, while this is not the case if we include the rotationally corrected size of the core. It is thus crucial to include the effects of rotation in the bulk viscosity calculation. This was already noted by \cite{Nayyar}.
Note that the effect of rotation on the size of the core is technically an order $\mathcal{O}(\Omega^2)$ effect compared to the leading order term (which would give the extent of the core in a non rotating star). This means that, to be consistent when calculating the bulk viscosity dissipation integral we would have to consider higher order terms also in the mode eigenfunctions, i.e. compute the whole integral to order $\mathcal{O}(\Omega^4)$. This is, of course, prohibitive. We do, however, feel that it is appropriate to calculate the bulk viscosity damping for a sequence of stars with different rotation rates and that, when the core becomes very small (i.e. for large rotation rates) the effect of the reduced size will dominate (in other words we assume that the eigenfunctions of the mode are regular and that they will not contain terms that, at second order in rotation, become very large as the core becomes very small). It is important to bear in mind, however, that we are neglecting terms that could be of the same order as the change in size of the core and that could, especially for low rotation rates, lead to quantitative (but probably not qualitative) differences in the damping timescale.

\begin{figure}
\centerline{\includegraphics[height=7cm,clip]{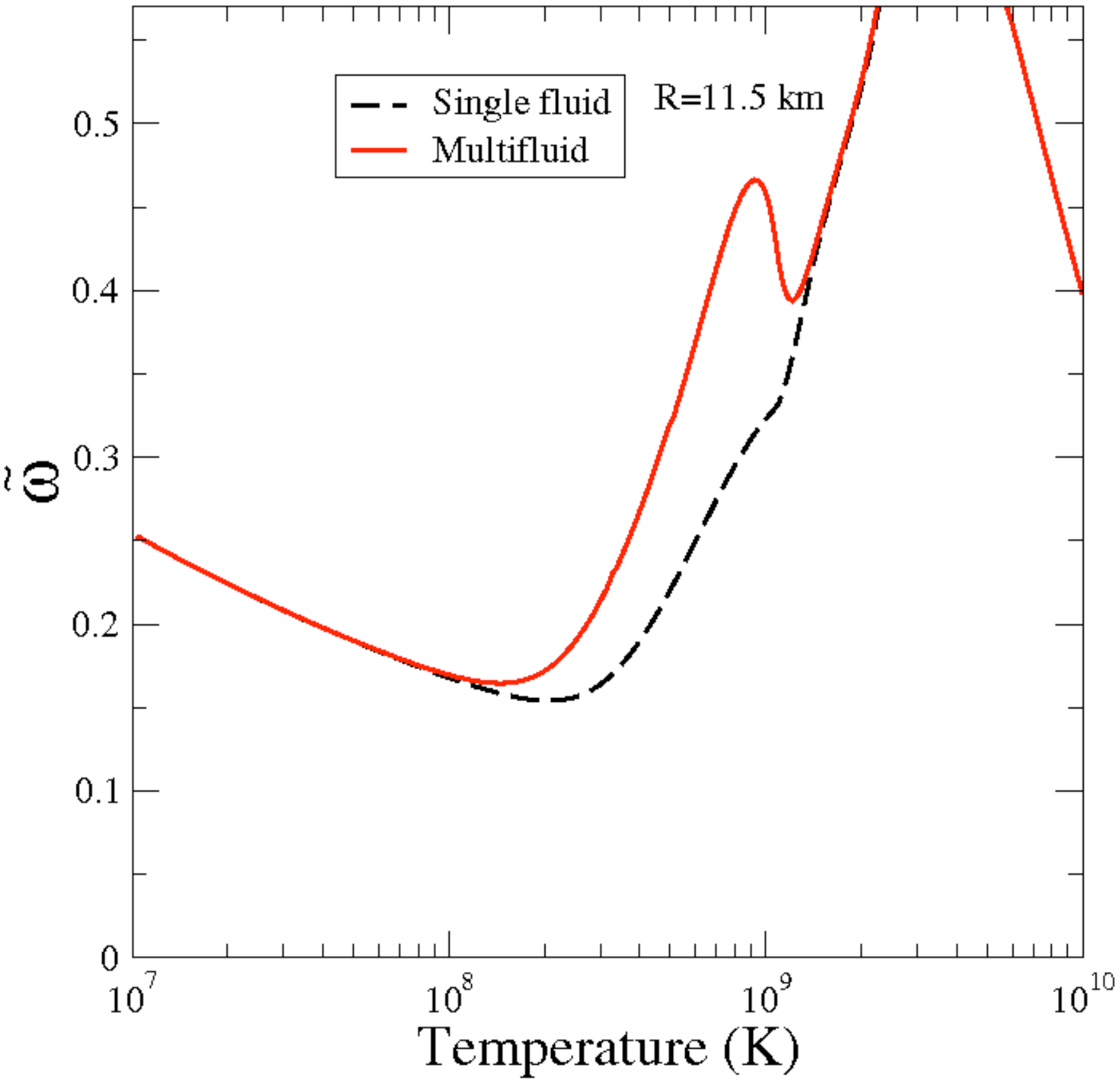}\includegraphics[height=7cm,clip]{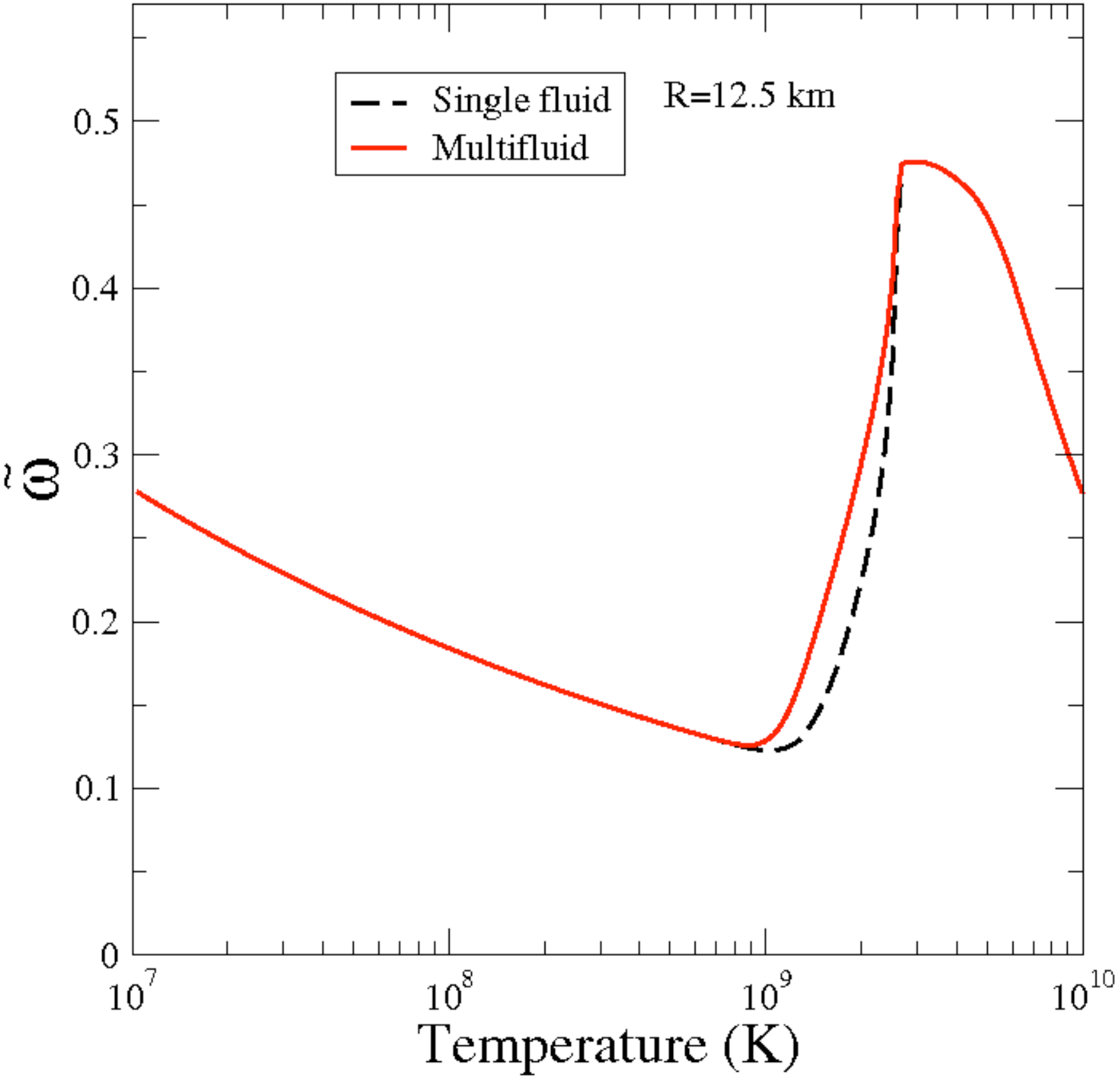}}
\centerline{\includegraphics[height=7cm,clip]{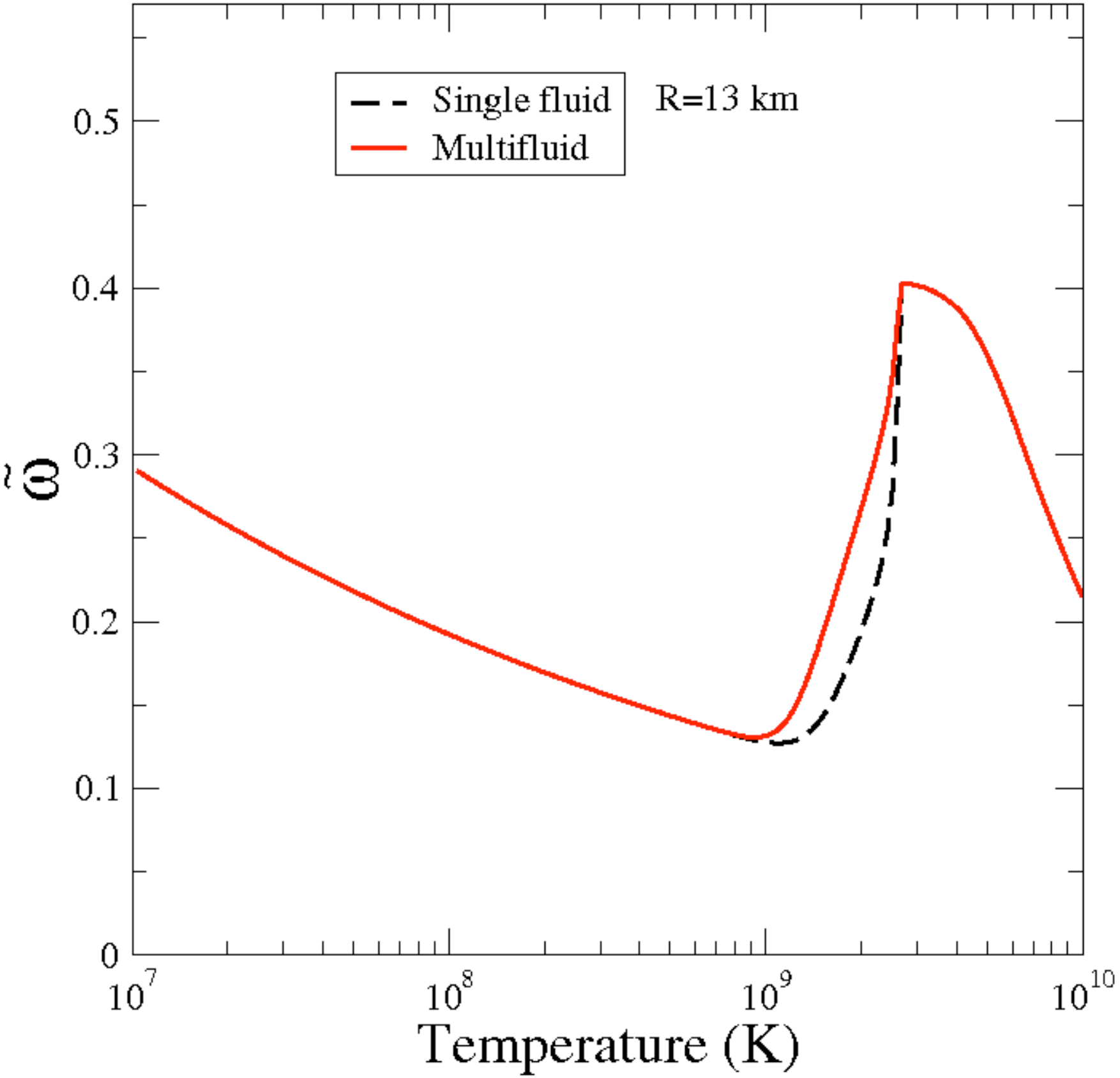}\includegraphics[height=7cm,clip]{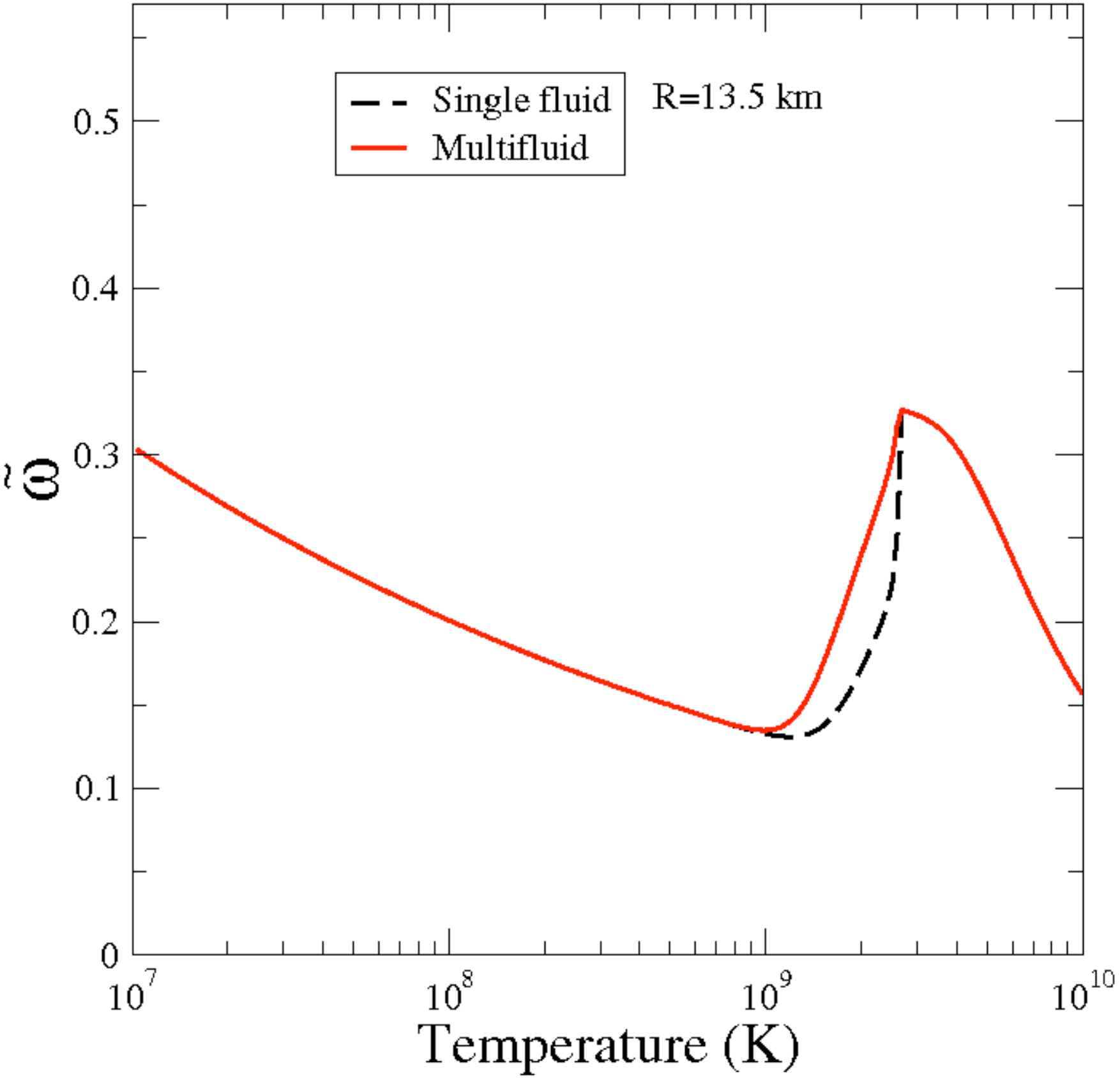}}
\caption{R-mode instability window, for a star with $M=1.4 M_{\odot}$ and varying radius. We show a comparison between the one-fluid case and the case with the two-fluid corrections included. The graphs extend to the breakup frequency and rotational corrections are included in both cases. The new "superfluid" coefficients increase the viscous damping but do not, in general, alter the qualitative features. }
\label{multi}
\end{figure}

\begin{figure}
\centerline{\includegraphics[height=7cm,clip]{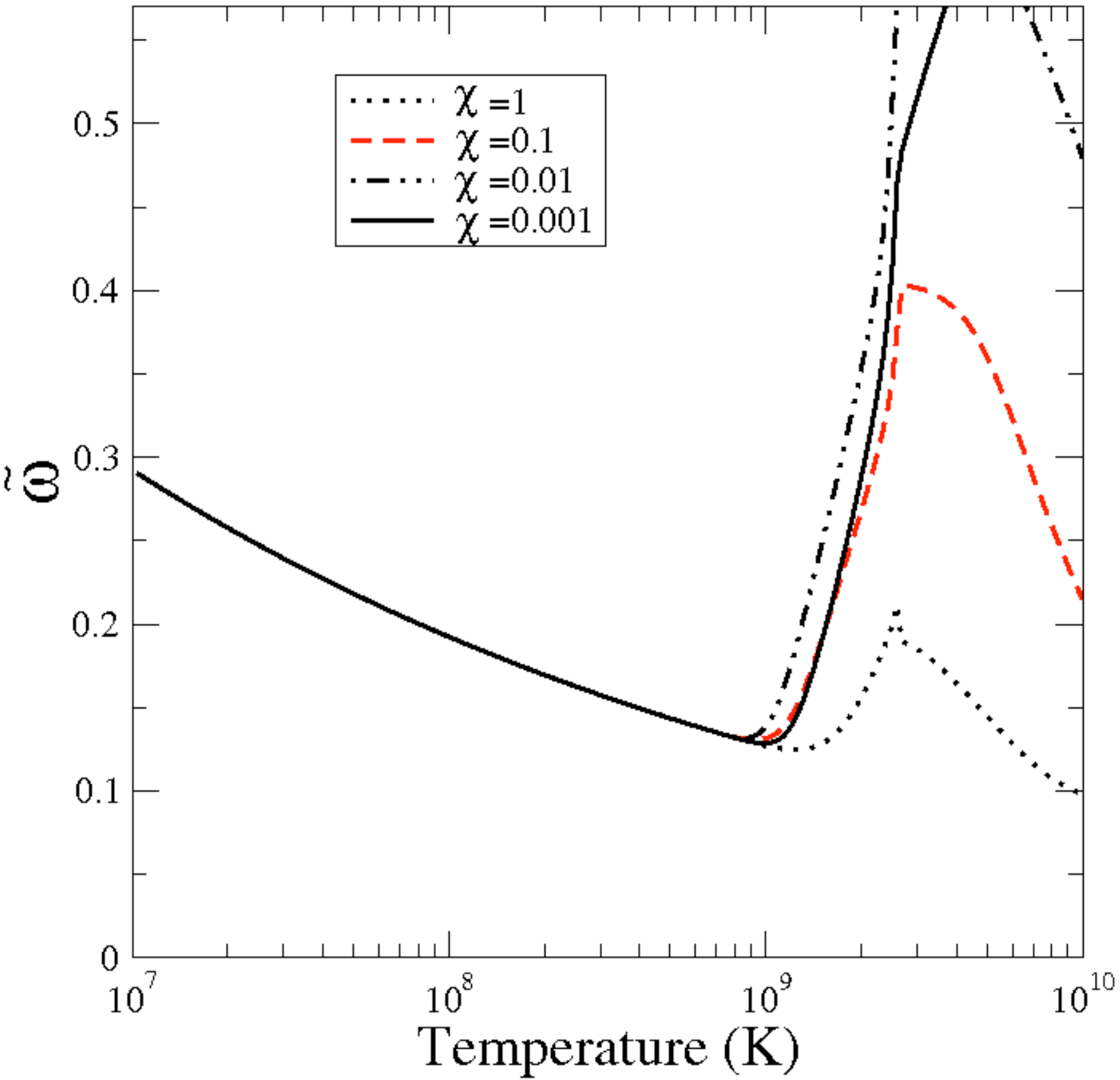}\includegraphics[height=7cm,clip]{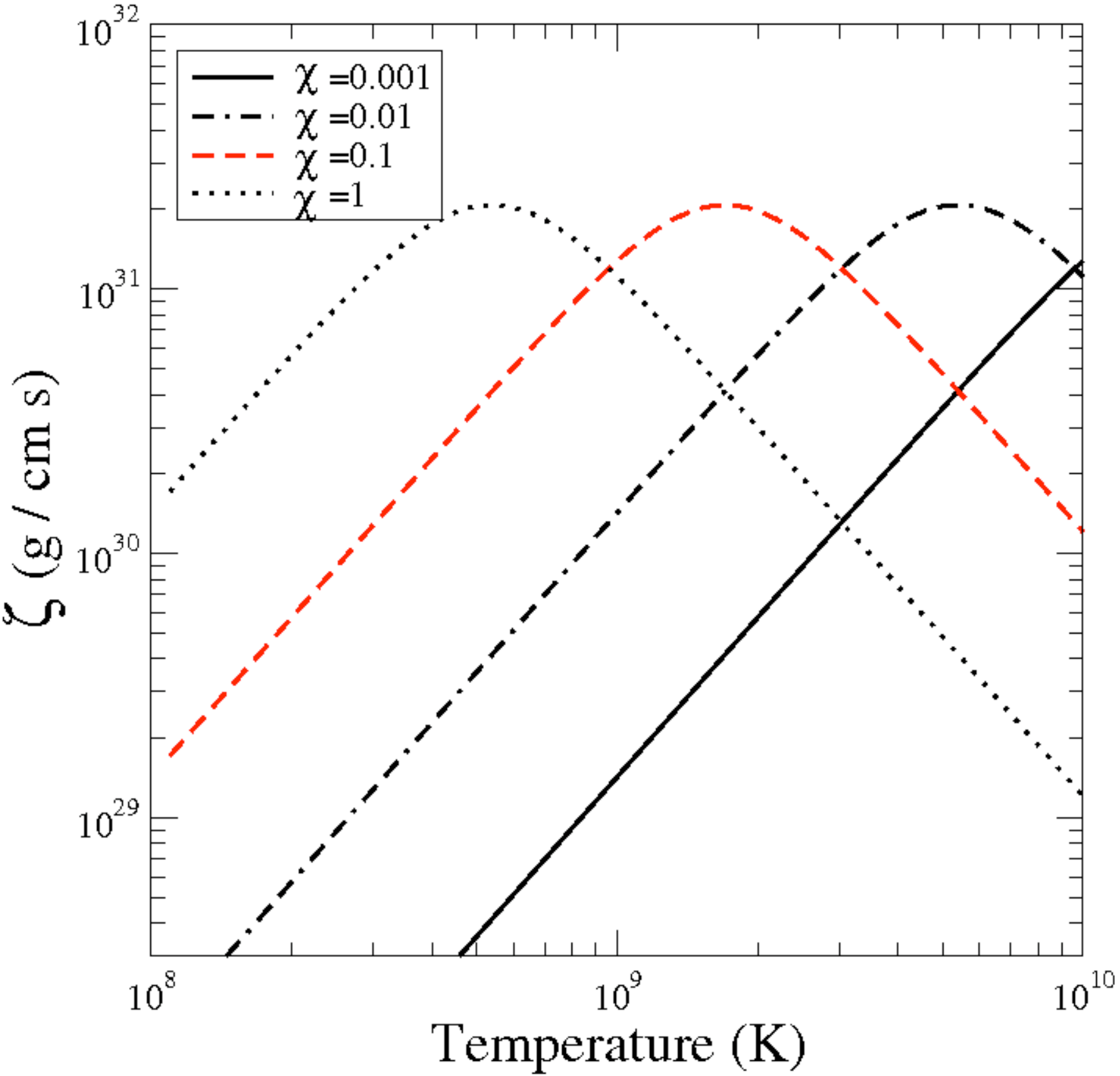}}
\caption{In the left panel we show the r-mode instability window for various values of the parameter $\chi$ in the hyperon bulk viscosity, for a $M=1.4 M_{\odot}$ and $R=13$ km star. The graphs extend to the breakup frequency. We assume that the viscosity at low temperatures is due to an Ekman layer at the base of the crust and that the reaction rates are reduced by superfluidity as prescribed in section \ref{super}. We see that varying $\chi$ can have an significant effect on the height of the resonance but not on the qualitative nature of the curve. On the right we show the bulk viscosity coefficient $\zeta$ as a function of temperature for $n_b$=1 fm$^{-3}$ and an oscillation frequency of 2300 Hz (chosen to directly compare with \citet{Nayyar}). One can easily appreciate that the maximum of the coefficient is shifted by changing $\chi$, leading to the differences seen in the left panel, as explained in the main text. }
\label{multi2}
\end{figure}

We can now consider the effect of introducing the extra ``superfluid'' bulk viscosity coefficients in the two-fluid region. In figure \ref{multi} we show the instability window for two stellar models, both in the one-fluid approximation and including the two-fluid fluid bulk viscosity coefficients, while in figure \ref{multi2} we show the effect of varying the stellar mass. Two effects are immediately obvious. First of all,  the effect of the extra coefficients is to increase the bulk viscosity damping and reduce the instability window. Although the effect is not very pronounced in general, it is clear from the left panel of figure \ref{multi} that, by increasing the strength of the damping in the low frequency region of the curve, multifluid effects could possibly suppress the instability in certain systems.  It is, however, also clear that the qualitative features of  the instability window are, in general, not affected and there is still a rise in the critical curve in the $T=10^9$ K region. This is due to the fact that the resonant nature of the bulk viscosity is robust.
The result was expected, since \cite{rmode} have shown that the r-mode is exclusively co-moving  at the leading order in rotation. This means that the main conclusion of \citet{strange} and \citet{Nayyar} is unaffected by the introduction of two-fluid effects. Notably, as the curve still has a positive slope in the $10^9$ K region, the possibility still exists that for certain systems hyperon bulk viscosity could halt the thermal runaway of the r-mode and lead to a neutron star  with a hyperon core becoming a persistent source of gravitational waves. Note, however, that if most of the hyperon core is not superfluid then the system may cool rapidly [see eg. \citet{page}]. In principle, this may halt a  thermal runaway  before the system reaches the positive sloping curve, leading once again to a limit cycle \citep{Bondarescu}.
The key point is that observations may be able to distinguish systems evolving according to the distinct scenarios, thereby providing insight into the qualitative nature of the r-mode 
instability curve, and by inference possibly information about the state of matter in the deep neutron star core.

Finally, before moving on, let us comment on the freedom associated with the parameter $\chi$. In the left panel of figure \ref{multi2} we illustrate the effect of varying this parameter in the range $0.001<\chi<1$. Clearly the effect on the height (frequency) of the peak of the resonance can be quite drastic, even though the qualitative nature of the curve is essentially unaffected, as it still exhibits a rise in the $T=10^9$ K region. However for large values of $\chi$  the effect of hyperon bulk viscosity is weakened, making it much less likely that the  thermal run-away of the system could be halted. We can gain an understanding of this from the right panel of figure \ref{multi2}, where we show the bulk viscosity coefficient $\zeta$ for varying $\chi$ at $n_b$=1 fm$^{-3}$. It is clear that larger values of $\chi$ shift the peak to lower temperatures, where bulk viscosity is not the main damping mechanism, and make the effect weaker in the higher temperature region. Such a large range of values for $\chi$ is, however, unlikely and we choose to use the fiducial value of $\chi=0.1$, as \citet{Gusakov2008} show that it reproduces their results and those of \citet{LO} to within factors of a few. It would obviously be good if future work were to better quantify the effects of the bare-particle assumption and narrow down the possible range for $\chi$.

\section{Concluding remarks}

Our results represent the first investigation into the effect of including the extra superfluid bulk viscosity coefficients in a calculation of the r-mode instability window. We have shown that, even though  the additional bulk viscosity coefficients do not alter the qualitative aspects of the instability window, there are regions of parameter space in which they could play a significant role, and may even suppress the instability entirely. In the light of these results we believe it is important to move beyond the qualitative analysis presented here. One should clearly account for the presence of $\Lambda$ hyperons and use a more realistic equation of state to describe the star. Furthermore, if one is to construct a more realistic model one clearly needs to work in general relativity and include finite temperature effects (such as in the entrainment coefficients calculated by \citet{Gusakov2}) and dissipative effects such as hyperon bulk viscosity. More detailed theoretical input is also needed from the nuclear physics community, in order to calculate the superfluid reaction rates needed to evaluate the bulk viscosity coefficients.

Developing the relevant tools will allow us to make progress on a range of related problems, e.g. involving finite temperature superfluid in the outer neutron star core or exotic phases of deconfined quarks in the deep core. An improved understanding of these systems is crucial if we are to take advantage of the unique opportunity that gravitational-wave detection could offer for the study of matter under extreme conditions.

\section*{Acknowledgments}

BH acknowledges support from the European Science Foundation (ESF) for the activity entitled 'The New Physics of Compact Stars', under exchange grant 2449, and thanks the Dipartimento di Fisica, Universit\`a degli studi di Milano for kind hospitality during part of this work.
We acknowledge support from STFC via grant number PP/E001025/1.

\appendix

\section{The entrainment matrix}

In this Appendix we  describe how to translate the results of \citet{Gusakov2009} into our formalism.
First of all we shall write the momentum of each fluid in the form $\pi_i^{\X}=\rho_{\X\Y} v_i^{\X}$, where the entrainment matrix $\rho_{\X\Y}$ encodes the fact that the momentum of each component is not necessarily parallel to it's velocity in a superfluid. For a system of $\n$, $\p$, $\Lambda$ and $\Sigma^{-}$ the momenta take the form:
\beq
\pi^i_\n&=&\rho^{\n\n}v^i_{\n}+\rho^{\n\p}v^i_{\p}+\rho^{\n\s}v^i_\s+\rho^{\n\l}v^i_\l\label{ma1}\\
\pi^i_\p&=&\rho^{\n\p}v^i_{\n}+\rho^{\p\p}v^i_{\p}+\rho^{\p\s}v^i_\s+\rho^{\p\l}v^i_\l\\
\pi^i_\s&=&\rho^{\n\s}v^i_{\n}+\rho^{\p\s}v^i_{\p}+\rho^{\s\s}v^i_\s+\rho^{\s\l}v^i_\l\\
\pi^i_\l&=&\rho^{\n\l}v^i_{\n}+\rho^{\p\l}v^i_{\p}+\rho^{\s\l}v^i_\s+\rho^{\l\l}v^i_\l\label{ma2}
\eeq
Were the entrainment matrix $\rho^{\X\Y}$ is connected to the relativistic entrainment matrix $Y^{\X\Y}$ of \citet{Gusakov2009} by
\be
Y^{\X\Y}=\frac{\rho^{\X\Y}}{c^2 m_\X m_\Y}
\ee
By comparing with the momenta in equations (\ref{m11})--(\ref{mom}) one the finds that:
\beq
c^2Y^{\n\n}&=&\frac{1}{m_\n^2}\left(m^\n n_\n-2\alpha^{\n\p}-2\alpha^{\n\s}-2\alpha^{\n\l}\right)\\
c^2Y^{\n\p}&=&\frac{2\alpha^{\n\p}}{m_\n m_\p},\;\;\;\;\;\;\;\;c^2 Y^{\n\s}=\frac{2\alpha^{\n\s}}{m_\n m_\s},\;\;\;\;\;\;\;\;c^2 Y^{\n\l}=\frac{2\alpha^{\n\l}}{m_\n m_\l}\\
c^2Y^{\p\p}&=&\frac{1}{m_\p^2}\left(m_\n n_\n-2\alpha^{\n\p}-2\alpha^{\p\s}-2\alpha^{\p\l}\right)\\
c^2Y^{\p\s}&=&\frac{2\alpha^{\p\s}}{m_\p m_\s},\;\;\;\;\;\;\;\;c^2 Y^{\p\l}=\frac{2\alpha^{\p\l}}{ m_\p m_\l}\,;\;\;\;\;\;\;\;c^2 Y^{\s\l}=\frac{2\alpha^{\s\l}}{m_\s m_\l}\\
c^2Y^{\s\s}&=&\frac{1}{m_\s^2}\left(m^\s n_\s-2\alpha^{\n\s}-2\alpha^{\p\s}-2\alpha^{\s\l}\right)\\
c^2Y^{\l\l}&=&\frac{1}{m_\l^2}\left(m^\l n_\l-2\alpha^{\s\l}-2\alpha^{\p\l}-2\alpha^{\n\l}\right)
\eeq
where the $\alpha^{\X\Y}$ are the entrainment parameters that enter the equations of motion in section 3.1.
We shall restrict ourselves to the case of no $\Lambda$ hyperons and assume that the components of the relativistic entrainment matrix are constant in the core. In particular, as an approximation to the results in figure 1 of \citet{Gusakov2009}, we shall assume that $Y^{\n\s}=Y^{\n\p}$, from which we can obtain that:
\be
\alpha^{\n\p}=\alpha^{\n\s}+\mathcal{O}(\Delta m/m_\n)
\ee
which leads to the result used to calculate the dissipation integral in (\ref{2fluid}):
\be
f(\bar{\varepsilon})=1+\mathcal{O}(\Delta m/m_\n)^2
\ee


\begin{thebibliography}{99}
\bibitem[\protect\citeauthoryear{Alpar, Langer \& Sauls}{Alpar, Langer \& Sauls}{1984}]{ALS} Alpar M.A., Langer S.A., Sauls J.A. 1984, ApJ, 282, 533
\bibitem[\protect\citeauthoryear{Andersson}{Andersson}{2003}]{anderssonREVIEW} Andersson N., 2003, CQG, 20, R105
\bibitem[\protect\citeauthoryear{Andersson \& Comer}{Andersson \& Comer}{2001}]{ac2001} Andersson N., Comer G.L., 2001, MNRAS, 328, 1129
\bibitem[\protect\citeauthoryear{Andersson, Comer \& Glampedakis}{Andersson et al.}{2005}]{nuclphys} Andersson N., Comer G.L., Glampedakis K.,2005, Nucl. Phys. A, 763, 212
\bibitem[\protect\citeauthoryear{Andersson \& Comer}{Andersson \& Comer}{2006}]{CQG} Andersson N., Comer G.L., 2006, CQG, 23, 5503
\bibitem[\protect\citeauthoryear{Andersson \& Comer}{Andersson \& Comer}{2008}]{helium} Andersson N., Comer G.L., 2008, eprint arXiv:0811.1660
\bibitem[\protect\citeauthoryear{Andersson, Glampedakis \& Samuelsson}{Andersson et al.}{2010}]{MHD} Andersson N., Glampedakis K., Samuelsson L., 2010, eprint arXiv:1001.4046
\bibitem[\protect\citeauthoryear{Andersson, Jones \& Kokkotas}{Andersson et al.}{2002}]{strange} Andersson N., Jones, D.I., Kokkotas K.D., 2002, MNRAS, 337, 1224
\bibitem[\protect\citeauthoryear{Andersson \& Kokkotas}{Andersson \& Kokkotas}{2001}]{review} Andersson N., Kokkotas K.D., 2001, Int. J. Mod. Phys. D, 10, 381
\bibitem[\protect\citeauthoryear{Andersson, Kokkotas \& Stergioulas}{Andersson et al.}{1999}]{AKS} Andersson N., Kokkotas K.D., Stergioulas N., 1999, ApJ., 516, 307
\bibitem[\protect\citeauthoryear{Andersson, Sidery \& Comer}{Andersson et al.}{2006}]{Trev} Andersson N., Sidery T., Comer G.L., 2006, MNRAS, 368, 162
\bibitem[\protect\citeauthoryear{Arras et al.}{Arras et al.}{2003}]{Arras} Arras P. et al., 2003, ApJ, 591, 1129
\bibitem[\protect\citeauthoryear{Balberg \& Barnea}{Balberg \& Barnea}{1998}]{BB} Balberg S., Barnea N., 1998, Phys. Rev. C, 57, 409
\bibitem[\protect\citeauthoryear{Bildsten}{Bildsten}{1998}]{Bildsten} Bildsten L., 1998, ApJ Letters, 501, L89
\bibitem[\protect\citeauthoryear{Bildsten \& Ushomirsky}{Bildsten \& Ushomirsky}{2000}]{BU} Bildsten L., Ushomirsky G., 2000, ApJ, 529, L33
\bibitem[\protect\citeauthoryear{Bondarescu, Teukolsky \& Wasserman}{Bondarescu et al.}{2009}]{Bondarescu} Bondarescu R., Teukolsky S.A.. Wasserman I., 2009, Phys. Rev. D, 79, 104003
\bibitem[\protect\citeauthoryear{Brink, Teukolsky \& Wasserman}{Brink et al.}{2004}]{Brink} Brink J., Teukolsky S.A.. Wasserman I., 2004, Phys. Rev. D, 70, 121501
\bibitem[\protect\citeauthoryear{Chandrasekhar}{Chandrasekhar}{1933}]{Chandra} Chandrasekhar S., 1933, MNRAS, 93, 390
\bibitem[\protect\citeauthoryear{Epstein}{Epstein}{1988}]{Epstein} Epstein R., 1988, ApJ, 333, 880
\bibitem[\protect\citeauthoryear{Gaertig \& Kokkotas}{Gaertig \& Kokkotas}{2008}]{Gar} Gaertig E.., Kokkotas K.D., 2008, Phys. Rev. D, 78, 064063
\bibitem[\protect\citeauthoryear{Glampedakis \& Andersson}{Glampedakis \& Andersson}{2006a}]{eck1} Glampedakis K., Andersson N., 2006a, MNRAS, 371, 1311
\bibitem[\protect\citeauthoryear{Glampedakis \& Andersson}{Glampedakis \& Andersson}{2006b}]{eck2} Glampedakis K., Andersson N., 2006b, Phys. Rev. D, 74, 044040
\bibitem[\protect\citeauthoryear{Glendenning}{Glendenning}{1985}]{Glen} Glendenning N.K., 1985, ApJ, 293, 470
\bibitem[\protect\citeauthoryear{Glendenning}{Glendenning}{1996}]{Glenbook} Glendenning N.K., "Compact Stars", 1996, Springer-Verlag, New York
\bibitem[\protect\citeauthoryear{Gusakov \& Kantor}{Gusakov \& Kantor}{2008}]{Gusakov2008} Gusakov M.E., Kantor E.M., 2008, Phys. Rev. D, 78, 83006
\bibitem[\protect\citeauthoryear{Gusakov, Kantor \& Haensel}{Gusakov et al.}{2009a}]{Gusakov2009} Gusakov M.E., Kantor, E.M., Haensel P., 2009, Phys. Rev. C, 79, 55806
\bibitem[\protect\citeauthoryear{Gusakov, Kantor \& Haensel}{Gusakov et al}{2009b}]{Gusakov2} Gusakov M.E., Kantor, E.M., Haensel P., 2009, Phys. Rev. C, 80, 015803
\bibitem[\protect\citeauthoryear{Haskell, Andersson \& Passamonti}{Haskell et al.}{2009}]{rmode} Haskell B., Andersson N., Passamonti A., 2009, MNRAS, 397, 1464
\bibitem[\protect\citeauthoryear{Haskell, Andersson \& Comer}{Haskell et al.}{(in preparation)}]{prep} Haskell B., Andersson N., Comer G.L., 2010, in preparation
\bibitem[\protect\citeauthoryear{Haensel, Levenfish \& Yakovlev}{Haensel et al.}{2002}]{haensel} Haensel P., Levenfish K.P., Yakovlev D.G., 2002, A\&A, 381, 1080
\bibitem[\protect\citeauthoryear{Heyl}{Heyl}{2002}]{Heyl} Heyl J., 2002, ApJ, 574, L57
\bibitem[\protect\citeauthoryear{Jones}{Jones}{2001}]{PBJones} Jones P.B., 2001, Phys.Rev. D, 64, 084003
\bibitem[\protect\citeauthoryear{Kantor \& Gusakov}{Kantor \& Gusakov}{2009}]{Gusakovsound} Kantor E.M., Gusakov M.E., 2009, Phys. Rev. D, 79, 43004
\bibitem[\protect\citeauthoryear{Kinney \& Mendell}{Kinney \& Mendell}{2003}]{Mend} Kinney, J.B., Mendell G., 2003, Phys. Rev. D, 67, 024032
\bibitem[\protect\citeauthoryear{Lattimer \& Prakash}{Lattimer \& Prakash}{2006}]{lp} Lattimer J.M., Prakash M., 2006, Nucl. Phys. A, 777, 479
\bibitem[\protect\citeauthoryear{Lee \& Yoshida}{Lee \& Yoshida}{2003}]{LY} Lee U., Yoshida S., 2003, ApJ, 586, 403
\bibitem[\protect\citeauthoryear{Levin}{Levin}{1999}]{levin1} Levin Y., 1999, ApJ, 517, 328
\bibitem[\protect\citeauthoryear{Levin \& Ushomirsky}{Levin \& Ushomirsky}{2001}]{LU} Levin Y., Ushomirsky, G., 2001, ApJ, 324, 917
\bibitem[\protect\citeauthoryear{Lindblom \& Mendell}{Lindblom \& Mendell}{1994}]{LM94} Lindblom L., Mendell G., 1994, ApJ, 321, 689
\bibitem[\protect\citeauthoryear{Lindblom \& Mendell}{Lindblom \& Mendell}{2000}]{LM} Lindblom L., Mendell G., 2000, Phys. Rev. D, 61, 104003
\bibitem[\protect\citeauthoryear{Lindblom \& Owen}{Lindblom \& Owen}{2002}]{LO} Lindblom L., Owen B.J., 2002, Phys. Rev. D, 65, 63006
\bibitem[\protect\citeauthoryear{Lindblom Owen \& Ushomirsky}{Lindblom et al.}{2000}]{LOU} Lindblom L., Owen B.J., Ushomirsky G., 2000, Phys. Rev. D, 62, 084030
\bibitem[\protect\citeauthoryear{Lockitch, Andersson \& Friedman}{Lockitch et al.}{2001}]{Lockitch1} Lockitch K.H., Andersson N., Friedman J.L., 2001, Phys. Rev. D, 63, 024019
\bibitem[\protect\citeauthoryear{Lockitch, Friedman \& Andersson}{Lockitch et al.}{2003}]{Lockitch} Lockitch K.H., Friedman J.L., Andersson N., 2003, Phys. Rev. D, 68, 124010
\bibitem[\protect\citeauthoryear{Mendell}{Mendell}{1991a}]{Ma} Mendell G., 1991, ApJ, 380, 515
\bibitem[\protect\citeauthoryear{Mendell}{Mendell}{1991b}]{Mb}  Mendell G., 1991, ApJ, 380. 530
\bibitem[\protect\citeauthoryear{Nayyar \& Owen}{Nayyar \& Owen}{2006}]{Nayyar} Nayyar B., Owen B.J., 2006, Phys. Rev. D, 73, 084001
\bibitem[\protect\citeauthoryear{\"Ozel, Baym \& G\"uver}{\"Ozel, Baym \& G\"uver}{2010}]{Ozel} \"Ozel. F, Baym G., G\"uver, T., 2010, preprint: arXiv:1002.3153v1
\bibitem[\protect\citeauthoryear{Owen, Lindblom, Cutler, Schutz, Vecchio \& Andersson}{Owen et al.}{1998}]{O98} Owen B.J., Lindblom L., Cutler C., Schutz B.F., Vecchio, A., Andersson N., 1998, Phys. Rev. D, 58, 084020
\bibitem[\protect\citeauthoryear{Page, Geppert \& Weber}{Page, Geppert \& Weber}{2006}]{page} Page D., Geppert U., Weber F., 2006, Nucl. Phys. A, 777, 497
\bibitem[\protect\citeauthoryear{Pons, Gualtieri, Miralles \& Ferrari}{Pons et al.}{2005}]{jose} Pons J.A., Gualtieri L., Miralles J.A., Ferrari V., 2005, MNRAS, 363, 121
\bibitem[\protect\citeauthoryear{Ruoff \& Kokkotas}{Ruoff \& Kokkotas}{2001}]{Ruoff01} Ruoff J., Kokkotas K.D., 2001, MNRAS, 328, 678
\bibitem[\protect\citeauthoryear{Ruoff \& Kokkotas}{Ruoff \& Kokkotas}{2002}]{Ruoff02} Ruoff J., Kokkotas K.D., 2002 MNRAS, 330, 1027
\bibitem[\protect\citeauthoryear{Sawyer}{Sawyer}{1989}]{Sawyer1989} Sawyer R.F., 1989, Phys. Rev. D, 39, 3804
\bibitem[\protect\citeauthoryear{Sidery}{Sidery}{2008}]{TThesis} Sidery T., 2008, PhD Thesis, University of Southampton
\bibitem[\protect\citeauthoryear{Yoshida, Yoshida \& Eriguchi}{Yoshida et al.}{2005}]{Yoshida} Yoshida S., Yoshida S., Eriguchi Y., 2005 MNRAS, 356, 217
\bibitem[\protect\citeauthoryear{Wagoner}{Wagoner}{2004}]{wagoner2004} Wagoner R.V., 2004, "X-ray timing 2003:Rossie and Beyond. AIP Conference Proceedings", 714, 224
\bibitem[\protect\citeauthoryear{Watts, Krishnan, Bildsten \& Schutz}{Watts et al.}{2008}]{watts2008} Watts A.L., Krishnan B., Bildsten L., Schutz B.F., 2008, MNRAS, 389, 839
\end{thebibliography}
\end{document}